\documentclass[twocolumn]{aastex63}

\received{TBD}
\accepted{TBD}
\acceptjournal{ApJ}

\submitjournal{ApJ}

\shorttitle{Burstiness at $z=3$--9}
\shortauthors{Mitsuhashi et al.}

\graphicspath{{./}}
\renewcommand\baselinestretch{0.92}
\usepackage{amsmath}
\tabletypesize{\footnotesize}
\usepackage{multirow}

\usepackage{ulem} 
\def\blue#1 {{\textcolor{blue}{#1}}\ }

\def\oiii{[O\,{\sc iii}]}

\def\hanii{H$\alpha$+[N\,{\sc ii}]}
\def\oiiihb{[O\,{\sc iii}]+H$\beta$}
\def\kms{\,km\,s$^{-1}$}

\def\red#1 {{\textcolor{red}{#1}}\ }

\def\rha{$R_{{\rm H}\alpha}$}
\def\roiii{$R_{{\rm [OIII]}}$}
\def\logrha{$\log\!R_{{\rm H}\alpha}$}
\def\logroiii{$\log\!R_{\rm [OIII]}$}

\def\ewha{${\rm EW}_{{\rm H}\alpha}$}
\def\ewoiii{${\rm EW}_{{\rm [O\,III]}}$}
\def\ppha{$P^{84-16}_{\log\!R_{{\rm H}\alpha}}$}
\def\ppoiii{$P^{84-16}_{\log\!R_{\text{[O\,{\sc iii}]}}}$}

\def\logroiiihb{$\log\!R_{{\rm [OIII]}+{\rm H}\beta}$}
\def\rline{$R_{\rm line}$}

\def\ewhanii{${\rm EW}_{{\rm H}\alpha+[{\rm N\,II}]}$}

\begin{document}


\title{UNCOVER/MegaScience Finds Uniform and Highly Bursty Star Formation at $3<z<9$, \\
consistent with the High-Redshift UV Luminosity Function}

\correspondingauthor{Ikki Mitsuhashi}
\email{ikki0913astr@gmail.com}

\author[0000-0001-7300-9450]{Ikki Mitsuhashi}
\affiliation{Department for Astrophysical \& Planetary Science, University of Colorado, Boulder, CO 80309, USA}

\author[0000-0002-1714-1905]{Katherine A. Suess}
\affiliation{Department for Astrophysical \& Planetary Science, University of Colorado, Boulder, CO 80309, USA}

\author[0000-0001-6755-1315]{Joel Leja}
\affiliation{Department of Astronomy \& Astrophysics, The Pennsylvania State University, University Park, PA 16802, USA}
\affiliation{Institute for Computational \& Data Sciences, The Pennsylvania State University, University Park, PA 16802, USA}
\affiliation{Institute for Gravitation and the Cosmos, The Pennsylvania State University, University Park, PA 16802, USA}

\author[0000-0001-5063-8254]{Rachel Bezanson}
\affiliation{Department of Physics and Astronomy and PITT PACC, University of Pittsburgh, Pittsburgh, PA 15260, USA}

\author[0000-0002-5612-3427]{Jenny E. Greene}
\affiliation{Department of Astrophysical Sciences, Princeton University, Princeton, NJ 08544, USA}

\author[0000-0001-8174-317X]{Emilie Burnham}
\affiliation{Department of Astronomy \& Astrophysics, The Pennsylvania State University, University Park, PA 16802, USA}
\affiliation{Institute for Gravitation and the Cosmos, The Pennsylvania State University, University Park, PA 16802, USA}
\affiliation{Center for Astrostatistics and Astroinformatics, The Pennsylvania State University, University Park, PA 16802, USA}

\author[0000-0002-3475-7648]{Gourav Khullar}
\affiliation{Department of Astronomy, University of Washington, Physics-Astronomy Building, Box 351580, Seattle, WA 98195-1700, USA}
\affiliation{Science Institute, University of Washington, Physics-Astronomy Building, Box 351580, Seattle, WA 98195-1700, USA}
\affiliation{Department of Physics and Astronomy and PITT PACC, University of Pittsburgh, Pittsburgh, PA 15260, USA}

\author[0000-0002-9816-9300]{Abby Mintz}
\affiliation{Department of Astrophysical Sciences, Princeton University, Princeton, NJ 08544, USA}

\author[0000-0003-2804-0648 ]{Themiya Nanayakkara}
\affiliation{Centre for Astrophysics and Supercomputing, Swinburne University of Technology, PO Box 218, Hawthorn, VIC 3122, Australia}

\author[0000-0002-3254-9044]{Karl Glazebrook}
\affiliation{Centre for Astrophysics and Supercomputing, Swinburne University of Technology, PO Box 218, Hawthorn, VIC 3122, Australia}

\author[0000-0002-0108-4176]{Sedona H. Price}
\affiliation{Space Telescope Science Institute (STScI), 3700 San Martin Drive, Baltimore, MD 21218, USA}   
\affiliation{Department of Physics and Astronomy and PITT PACC, University of Pittsburgh, Pittsburgh, PA 15260, USA}  

\author[0000-0003-4075-7393]{David J. Setton}\thanks{Brinson Prize Fellow}
\affiliation{Department of Astrophysical Sciences, Princeton University, Princeton, NJ 08544, USA}

\author[0000-0001-9269-5046]{Bingjie Wang
}
\altaffiliation{NHFP Hubble Fellow}
\affiliation{Department of Astrophysical Sciences, Princeton University, Princeton, NJ 08544, USA}

\author[0000-0003-1614-196X]{John R. Weaver}\thanks{Brinson Prize Fellow}
\affiliation{MIT Kavli Institute for Astrophysics and Space Research, 70 Vassar Street, Cambridge, MA 02139, USA}

\author[0000-0002-7570-0824]{Hakim Atek}
\affiliation{Institut d’Astrophysique de Paris, CNRS, Sorbonne Universit\'e, 98bis Boulevard Arago, 75014, Paris, France}

\author[0000-0001-8460-1564]{Pratika Dayal}
\affiliation{Kapteyn Astronomical Institute, University of Groningen, 9700 AV Groningen, The Netherlands}
\affiliation{Canadian Institute for Theoretical Astrophysics, 60 St George St, University of Toronto, Toronto, ON M5S 3H8, Canada}
\affiliation{David A. Dunlap Department of Astronomy and Astrophysics, University of Toronto, 50 St George St, Toronto ON M5S 3H4, Canada}
\affiliation{Department of Physics, 60 St George St, University of Toronto, Toronto, ON M5S 3H8, Canada}

\author[0000-0002-1109-1919]{Robert Feldmann}
\affiliation{Department of Astrophysics, University of Zurich, CH-8057, Switzerland}

\author[0000-0001-7201-5066]{Seiji Fujimoto}
\altaffiliation{NHFP Hubble Fellow}
\affiliation{David A. Dunlap Department of Astronomy and Astrophysics, University of Toronto, 50 St George St, Toronto ON M5S 3H4, Canada}
\affiliation{Department of Astronomy, The University of Texas at Austin, Austin, TX 78712, USA}

\author[0000-0001-6278-032X]{Lukas J. Furtak}\affiliation{Physics Department, Ben-Gurion University of the Negev, P.O. Box 653, Be’er-Sheva 84105, Israel}

\author[0000-0002-5337-5856]{Brian Lorenz}
\affiliation{Department for Astrophysical \& Planetary Science, University of Colorado, Boulder, CO 80309, USA}
\affiliation{Department of Astronomy, University of California, Berkeley, CA 94720, USA}

\author{Natalia Porraz Barrera}
\affiliation{Department for Astrophysical \& Planetary Science, University of Colorado, Boulder, CO 80309, USA}

\author[0000-0002-2057-5376]{Ivo Labbe}
\affiliation{Centre for Astrophysics and Supercomputing, Swinburne University of Technology, Melbourne, VIC 3122, Australia}

\author[0000-0003-2680-005X]{Gabriel Brammer}
\affiliation{Cosmic Dawn Center (DAWN), Niels Bohr Institute, University of Copenhagen, Jagtvej 128, K{\o}benhavn N, DK-2200, Denmark}

\author[0000-0002-7031-2865]{Sam E. Cutler}
\affiliation{Department of Astronomy, University of Massachusetts, Amherst, MA 01003, USA}

\author[0000-0002-9651-5716]{Richard Pan}\affiliation{Department of Physics and Astronomy, Tufts University, 574 Boston Ave., Medford, MA 02155, USA}

\author[0000-0001-7160-3632]{Katherine E. Whitaker}
\affiliation{Department of Astronomy, University of Massachusetts, Amherst, MA 01003, USA}
\affiliation{Cosmic Dawn Center (DAWN), Denmark}

\author{the UNCOVER/MegaScience team}

\begin{abstract}
Star formation timescales are key to understanding fundamental physics like feedback mechanisms, as well as the abundance of bright galaxies at $z>10$.
We investigate galaxy star formation histories (SFHs) and their evolution across $z\sim3$--9 by measuring the line-to-UV ratio (\rline) and line equivalent width (EW) of \hanii\ and \oiiihb\ directly from UNCOVER/MegaScience spectro-photometry without relying on a specific SFH or nebular line modeling.
Our photometric measurements recover \rline\ and EW to $<10\%$ systematic accuracy compared to spectroscopy.
This allows us to construct a large mass- (and flux-) complete sample and quantitatively examine how \rline\ evolves with redshift and stellar mass.
We find that the intrinsic scatter in \rline\ does not significantly evolve with redshift across $3<z<7$, though it may increase at $z\gtrsim8$. 
We build population-level toy models using \texttt{fsps} to help interpret our observations, and find that scatter in \rline\ primarily reflects the amplitude of SFH fluctuations; this implies that our observed lack of evolution in the scatter of \rline\ is due to similar star formation burstiness from $z\sim3$ to $z\sim7$. 
Our observations are best reproduced by a set of SFHs with rising, long-duration, and large-amplitude bursts.
Finally, we demonstrate that the toy model that best describes our $z\sim6$ data can boost UV brightness by up to $\Delta M_{\rm UV}\sim-2.0\,{\rm mag}$ compared with a 200\,Myr constant SFH, and naturally produces a large number of galaxies at $z>10$. 
This suggests that no significant evolution in star formation burstiness is required to explain the abundance of UV-bright galaxies at high redshift.
\end{abstract}




%
%
%
%
%
%
\section{Introduction}

Constraining star formation timescales across the Universe is critical to shed light on the early galaxy formation models as well as the underlying physics, such as dominant feedback mechanisms \citep[e.g.,][]{1999MNRAS.310.1087S,2005MNRAS.361..776S,2009MNRAS.396.2332K,2018MNRAS.473.3717F}.
In addition to such fundamental aspects, bursty star formation histories (SFHs) have been proposed to explain the abundance of bright galaxies in the rest-frame ultra-violet (UV) at $z\gtrsim10$ revealed in recent JWST observations \citep[e.g.,][]{2022ApJ...938L..15C,2022ApJ...940L..55F,2022ApJ...940L..14N,2023MNRAS.518.6011D,2023Natur.616..266L,2023MNRAS.519.1201A,2023ApJS..265....5H,2023ApJ...957L..34W,2024ApJ...977..250F,2024ApJ...960...56H,2024Natur.633..318C,2024MNRAS.531.2615C,2024ApJ...972..143C,2025NatAs...9..155Z,2025A&A...693A..50N,2025arXiv250511263N}, as rapidly increasing star formation activities increase the fraction of hot O/B type stars and boost the visibility in the rest-frame UV \citep[e.g.,][]{2023MNRAS.525.2241H,2023MNRAS.526.2665S,2023ApJ...955L..35S,2023MNRAS.521..497M,2024arXiv240910613N}.
Additionally, identification of the so-called mini-quenching (or dormant, napping) galaxies, which halted their star formation in the recent 10\,Myrs, supported the necessity of the bursty SFHs \citep[e.g.,][]{2023ApJ...949L..23S,2024Natur.629...53L,2025A&A...697A..88L,2025MNRAS.537.3662T,2025arXiv250622540C}.

Several observable quantities trace star formation activity over different durations, such as emission lines (e.g., H$\alpha$, [O\,{\sc iii}], [O\,{\sc ii}]), UV continua, FIR continua, and Balmer break strengths \citep{1998ARA&A..36..189K,2014MNRAS.441.2717K}.
Among these observables, the relationship between the strength of the H$\alpha$ emission line and the far-UV continuum has traditionally been utilized as an indicator of star formation burstiness \citep[e.g.,][]{1999MNRAS.306..843G,2004A&A...421..887I,2007ApJ...671L.113L,2012ApJ...744...44W,2015MNRAS.451..839D,2019ApJ...881...71E}.
Far-UV continuum ($\lambda\sim1500\,$\AA) primarily originates from short-lived O-type stars as well as longer-lived B- and A-type stars, depending on the intrinsic recent SFHs.
For instance, sharply rising or falling SFH make O-type or A-type stars dominate the UV continuum, and therefore it traces the star-formation rate (SFR) within $>20$--$100\,{\rm Myrs}$ timescale \citep{2012ARA&A..50..531K}.
On the other hand, emission lines come from nebular regions ionized by O-type stars and reach equilibrium in $\sim5\,{\rm Myrs}$ in a constant SFH.
If constant star formation continues for $>100\,{\rm Myr}$, the luminosity ratio between H$\alpha$ and UV reaches an equilibrium value on $\gtrsim100\,{\rm Myr}$ timescale $-1.93\leq\text{\logrha}\leq-1.78$ \citep[][]{2023ApJ...952..133M}.
H$\alpha$/UV ratios higher than these equilibrium ratios indicate rising SFHs on $\sim5$--\,100Myr timescales, while lower H$\alpha$/UV ratios indicate falling SFHs.
Although other factors apart from SFH, such as the initial mass function (IMF), SPS models including binarity, ionizing photon leakage, stellar/gas phase metallicity, and dust, can also contribute to the variation in H$\alpha$-to-UV ratio \citep{2009ApJ...695..765M,2006ApJ...636..149E,2020ApJ...889..180N}, these variations are likely to be limited \citep[e.g., a $\sim10\%$ variation in the H$\alpha$-to-UV ratio or $\lesssim25\%$ of that induced by variations in the SFH,][]{2017ApJ...838...29M}. 

Spectroscopic observations are ideal for studying the H$\alpha$-to-UV ratio as they achieve accurate line and continuum measurements.
However, it is challenging to obtain a large number of spectra for a complete sample, even with the multiplexing capabilities of JWST/NIRSpec.
An alternative method is to infer the SFH of individual objects from spectral energy distribution (SED) fitting to their multi-band photometry.
Recent SED fitting tools introduce flexible time bins and/or SFRs and enable us to estimate the SFHs directly (e.g., \texttt{BEAGLE}; \citealt{2016MNRAS.462.1415C}, \texttt{Prospector}; \citealt{2019ApJ...877..140L,2021ApJS..254...22J}, and \texttt{Bagpipes}; \citealt{2018MNRAS.480.4379C}).
While these SED fitting tools are generally very useful, it is still challenging to accurately recover the SFHs of individual objects for several reasons.
First, stars formed in a recent burst have small mass-to-luminosity ratios ($M/L$), and outshine the older, larger $M/L$ stellar populations \citep[e.g.,][]{2001ApJ...559..620P}.
Second, the SFHs from SED fitting strongly depend on the prior and/or the choice of the time bins \citep[e.g.,][]{2022ApJ...927..170T,2022ApJ...935..146S,2025arXiv250415255W}.

One solution to break such difficulties is to derive SFHs at the population level: focusing on the distribution of the quantities that trace the SFH in a complete sample \citep[e.g.,][]{2025arXiv250415255W,2025arXiv250616510M,2024ApJ...961...53I}.
In this regard, imaging surveys with dense spectral sampling are the ideal dataset.
Dense spectral sampling allows us to accurately measure line and continuum fluxes.
Additionally, imaging surveys enable us to quantify the completeness at the given stellar mass or apparent fluxes - because the SFH itself affects the observability, careful sample selection is key for exploring burstiness of the SFH, and any biased dataset makes it unreliable \citep[e.g.,][]{2023MNRAS.526.2665S}.

Here we present a quantitative investigation of star formation timescales using the UNCOVER/MegaScience dataset, which includes all 20 broad- and medium-band filters on JWST/NIRCam.
This spectro-photometric dataset with spectral resolution of $R\!\sim\!15$ allows us to construct a stellar mass- (or flux-) complete sample without any biases introduced by sample selection in the spectroscopic follow-up.
We perform empirical line and continuum measurements, which are independent of physical stellar population synthesis models, and examine the scatter in the ratio between the emission line and continuum flux.
We then compare our observations with model SFHs at the population level instead of performing individual SED fitting.
In this study, we utilized both \hanii\ and \oiiihb, where H$\alpha$ and [N\,{\sc ii}] or [O\,{\sc iii}] and H$\beta$ are considered together due to line blending in $R\!\sim\!15$.
\hanii\ is a traditionally used, direct tracer of the SF timescale given a limited ($\lesssim10\%$) [N\,{\sc ii}] contribution in our target mass range of $\log M_{\ast}[M_{\odot}]=8$--9.5 at $z>3$ \citep{2023ApJS..269...33N,2023MNRAS.518..425C}.
While the interpretation of \oiiihb\ is more uncertain than \hanii\ due to its dependence on the ionizing condition, it is still useful to investigate SFHs given the similar large ionizing parameter $U_{\rm ion}$ and low gas-phase metallicity $Z$ in typical star-forming galaxies at $z>4$ \citep[e.g.,][]{2023ApJ...955...54S,2023ApJS..269...33N,2023MNRAS.518..425C}.
The strength of [O\,{\sc iii}] emission lines depends on other mechanisms apart from star formation, such as shocks and AGN activity \citep{2006MNRAS.372..961K,2008ApJS..178...20A}, but these rare phenomena are expected to have a limited effect on the whole population.

The paper is organized as follows: Section \ref{sec:data} provides an overview of the datasets used in this work. 
Section \ref{sec:analysis} describes the method of line and continuum measurements and model construction. 
In Section \ref{sec:results1}, we present our main results.
Discussion and comparison with previous studies are presented in Section \ref{sec:interp}. 
We finally summarized the paper in Section \ref{sec:summary}. 
Throughout this paper, we assume a flat universe with the cosmological parameters of $\Omega_{\rm M}=0.3$, $\Omega_{\Lambda}=0.7$, $\sigma_{8}=0.8$, and $H_0=70$ \kms ${\rm Mpc}^{-1}$.

%
%
%
%
%
%
\section{Data \& Sample}\label{sec:data}

\subsection{UNCOVER/MegaScience photometry}\label{subsec:uncover}
We utilize the JWST observations of the Abell 2744 cluster field from the UNCOVER \citep{2024ApJ...974...92B,2024ApJS..270....7W} and MegaScience \citep{2024ApJ...976..101S} surveys.
Specifically, we use the UNCOVER/MegaScience DR3 photometric catalog containing 20 NIRCam broad- and medium-band filters.
This catalog has includes data from several additional HST/JWST programs in the Abell 2744 field, such as GLASS-ERS \citep[PI: Treu][]{2022ApJ...935..110T}, ALT \citep[PI: Naidu \& Matthee][]{2024arXiv241001874N}, and MAGNIF \citep[PI: Sun][]{2023arXiv231009327L}, several GO programs (GO-3538, GO-2754; \citealt{2024MNRAS.528.7052C}), and HST surveys \citep[HFF and Buffalo,][]{2017ApJ...837...97L, 2020ApJS..247...64S}.
The DR3 catalog includes the physical properties of 74020 galaxies, such as the photometric redshift, the stellar mass, and dust extinction, based on the SED fitting results using the \texttt{Prospector}-$\beta$ model \citep{2023ApJ...944L..58W,2024ApJS..270...12W}.
Please refer to \citet{2024ApJ...976..101S} for more details on the DR4 catalog.

\subsection{Sample selection}\label{subsec:sample}
%
%
%
%
%
%
\begin{figure*}[htbp]
\begin{center}
\includegraphics[width=18cm,bb=0 0 1000 650, trim=0 1 0 0cm]{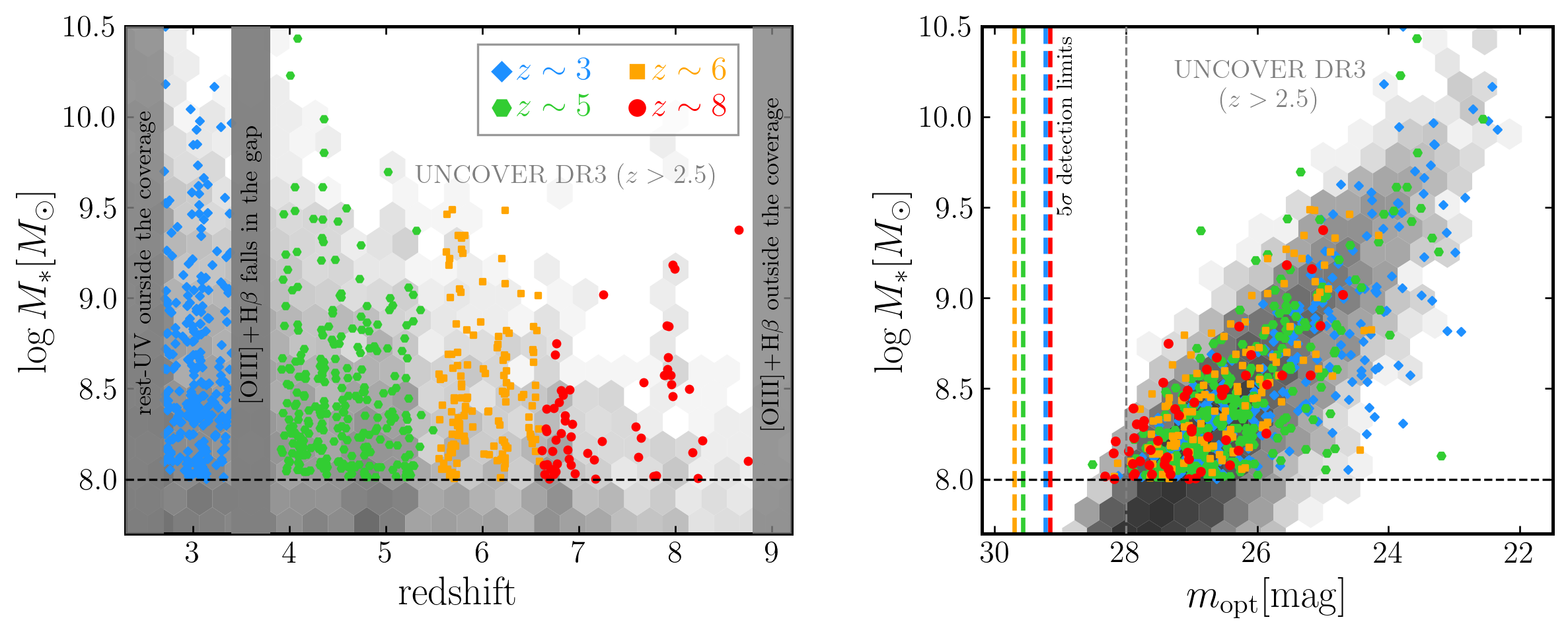}
\caption{(left) Sample selection for the mass-complete sample. The redshift gaps we avoided in the sample selection due to wavelength coverage are filled in gray. The background gray hexbins indicate all galaxies in the UNCOVER/MegaScience DR3 catalog. Our mass-limited samples satisfying $\log M_{\ast}[M_{\odot}]>8.0$ and some additional criteria (see text) at $z\sim3$, 5, 6, and 8 are shown in blue, green, orange, and red, respectively.
The different distributions between the selected sample and the parent UNCOVER DR3 sample arise from the criteria for the HST/medium band coverage or known AGN contribution.
(right) The relationship between the mass and rest-frame $\sim4000$\AA\ flux. Galaxies included in both mass-complete samples are shown in filled markers. The background gray hexbins indicate galaxies in the UNCOVER/MegaScience DR3 catalog at $z>2.5$. Typical $5\sigma$ detection limits in the filters tracing $m_{\rm opt}$, depending on the sample redshift (see text), are shown in the dashed lines with corresponding colors. The flux-complete sample is constructed from a threshold sufficiently smaller than $5\sigma$ detection limit in the UNCOVER survey (28\,mag). The mass-complete sample is mostly included in the flux-complete sample, enabling us to check whether the mass estimation from \texttt{Prospector}-$\beta$ affects the results significantly. }
\label{fig:sample}
\end{center}
\end{figure*}

From a parent sample of 74020 galaxies included in the DR3 catalog, we filtered out the artifacts and stars by requiring $\texttt{use\_phot}\!=\!1$ \citep{2024ApJS..270....7W}.
We then selected four subsamples, $z_{\rm best}=2.7$--3.5, 3.9--5.5, 5.5--6.6, and 6.6--8.8, where $z_{\rm best}$ is the redshift at which the redshift probability distribution $p(z)$ derived with \texttt{Prospector} is maximum.
We note that $p(z)$ recovers spectroscopic redshifts well with $\sigma_{\rm NMAD}\sim0.15$ and the $<10\%$ outlier fraction \citep[][]{2024ApJ...976..101S}.
Hereafter, we refer to these samples as $z\sim3$, 5, 6, and 8 samples, respectively. 
Our methods require that we cover both the rest-frame UV wavelengths and either the \hanii\ or \oiiihb\ emission lines. 
Therefore, we exclude $z=0$--2.7 because it lacks rest-frame UV coverage from either HST or JWST.
We additionally exclude galaxies at $z=3.5$--3.9 since [O\,{\sc iii}]+H$\beta$ falls a gap between the reddest SW band (F200W or F210M) and the bluest LW band (F277W or F250M, see Figures 2 and 6 in \citealt{2024ApJ...976..101S}).
Given NIRCam's coverage (0.7--5.0\,$\mu{\rm m}$), \hanii\ is measurable in our $z\sim3$, 5, 6 samples, and \oiiihb\ is measurable in our full sample.

For our $z\sim3$ and $z\sim5$ samples, we limit the sample to galaxies that fall inside the HST F435W and F606W images to cover the rest-frame UV wavelength.
We also exclude galaxies outside the full medium band coverage for the full samples to capture the line and continuum well.
We select galaxies with $\log_{10} M_{\ast}[M_{\odot}]>8.0$, above which 95\% of ``napping'' galaxies are detectable at the detection limit of the UNCOVER survey (see G. Khullar et al. in prep). 
The $M_{\rm UV}$ ranges $-23\lesssim M_{\rm UV}\,[{\rm mag}]\lesssim-16$ for $z\leq5$ samples and $-22\lesssim M_{\rm UV}\,[{\rm mag}]\lesssim-18$ for $z\geq5$ samples, respectively, which is sufficiently above the detection limit at the corresponding band ($M_{\rm UV}\leq-16\,{\rm mag}$).
Finally, we exclude known AGNs identified in \citet[][see also, \citealt{2024ApJ...964...39G}]{2025ApJ...978...92L}, which represent $<1\%$ of the sample.

The stellar masses used to construct the mass-limited sample are derived in \texttt{Prospector} SED fitting and depend on the SFH, which we will examine in this paper. 
To evaluate possible biases coming from the stellar mass estimation, we also constructed a flux-limited sample (Appendix \ref{appendix:fluxcomp}).
Here, we selected galaxies with apparent magnitudes at the rest-frame $\sim4000$\,\AA\ ($m_{\rm opt}$) brighter than 28.0\,AB\,mag. For the $z\sim3$, $z\sim5$, $z\sim6$, and $z\sim8$ samples, we used the F200W, F277W, F356W, and F444W filters, respectively, to estimate $m_{\rm opt}$.
At these filters, 28.0 mag is much brighter than 90\% complete flux \citep[by $\gtrsim1\,{\rm mag}$;][]{2024ApJ...974...92B,2024ApJS..270....7W}.

Our sample selection is summarized in Figure \ref{fig:sample}.
The mass-limited sample is almost contained in the flux-limited sample, making it possible to conservatively check whether the mass estimation from \texttt{Prospector}-$\beta$ would affect conclusions.
The total number of galaxies in the mass-(flux-) limited $z\sim3$, $z\sim5$, $z\sim6$, and $z\sim8$ samples are 258(536), 260(502), 130(264), and 58(99), respectively.

%
%
%
%
%
%
\section{Analysis}\label{sec:analysis}

\subsection{Photometric line/continuum measurements}\label{subsec:photmeasure}

%
%
%
%
%
%
\begin{figure*}[htbp]
\begin{center}
\includegraphics[width=18cm,bb=0 0 1000 650, trim=0 1 0 0cm]{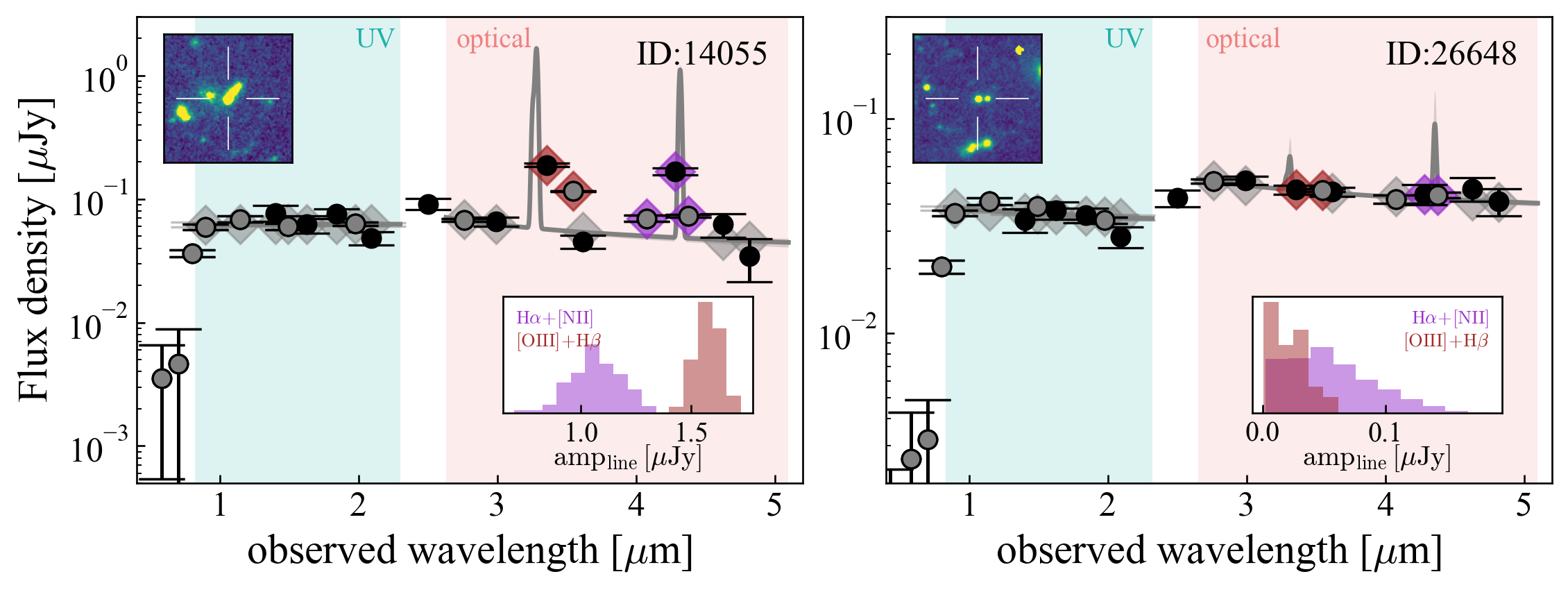}
\caption{Examples of our \texttt{emcee} fitting results for significant (left) and insignificant (right) line emitters.
Best-fit power-law(+Gaussian) model and its uncertainties are shown in gray lines and shaded regions, respectively. 
Blue and red wavelength ranges indicate the rest-frame UV and optical wavelengths specified in the fitting.
Observed photometries in broad+F410M and other medium bands are displayed in gray and black circles, respectively.
Gray diamonds show the best-fit photometry from the 50th percentile of the likelihood, overlaid in purple and brown for filters containing \hanii\ and \oiiihb, respectively. The inset panels at the bottom right in each panel show the posterior distribution of the \hanii\ (brown) and \oiiihb\ (purple) line amplitude in $\mu{\rm Jy}$ unit. The left top panels exhibit $2''\times2''$ thumbnails of the target galaxies.}
\label{fig:ex_fit}
\end{center}
\end{figure*}

We directly measure the rest-frame UV magnitude at $1500\,$\AA\ and the line flux of both \hanii\ and \oiiihb\ directly from NIRCam broad- and medium-band photometry, without any dependence on a physical stellar population synthesis model.
This is motivated to explore every parameter space at the given $z_{\rm phot}$, without any physical prior.

First, we identify rest-frame UV filters as those whose central wavelengths lie between 1250--3600\,$\text{\AA}\times(1+z_{\rm phot})$, and rest-frame optical filters as those between 4000--8500\,$\text{\AA}\times(1+z_{\rm phot})$.
We also define the filters containing emission lines [O\,{\sc iii}]\,$\lambda\lambda$4959,5007+H$\beta\,\lambda$4861 and H$\alpha\,\lambda6563$+[N\,{\sc ii}]\,$\lambda\lambda6548,6583$ as filters where the representative line wavelength (5007\AA\ for \oiiihb\ and 6563\AA\ for \hanii) is redshifted such that it lies at greater than 10\% of the maximum filter transmission.
Since \oiii\ and H$\beta$ are separated by  $\sim100\times(1+z)\,$\AA, more than one fifth of the effective width of some medium bands, \oiiihb\ is split into [O\,{\sc iii}] doublet and H$\beta$ if they are contained in different filter sets at the given $z_{\rm phot}$.
The aim of this split is not to measure \oiii\ and H$\beta$ independently, but to measure combined \oiiihb\ by summing them up.
\hanii\ concentrated within $\sim35\,$\AA\ range, so we do not split them.
Approximately 50\% of the sources fall within the redshift range at which \oiiihb\ is split into [O\,{\sc iii}] doublet and H$\beta$.

Next, we create simple model spectra, measure synthetic fluxes at each band using \texttt{sedpy}, and fit to the observed values using \texttt{emcee}.
The rest-frame UV spectrum is characterized by a power-law.
We set a log uniform prior with a range of [$10^{-5},10^{5}$] in $\mu{\rm Jy}$ for the amplitudes and a linear uniform prior with a range of [-5,5] for the slopes.
The rest-frame optical spectrum is represented by a combination of a power-law continuum and Gaussian emission lines.
We have two free parameters in the model continuum, amplitude and slope, and two (or three) parameters in the model line, the amplitude of blended \hanii\ and blended \oiiihb\ (or blended \hanii, blended [O\,{\sc iii}] doublet, and H$\beta$ when we split \oiiihb\ into [O\,{\sc iii}] and H$\beta$).
We fix the width of the Gaussian lines to $\sim500\,{\rm km}\,{\rm s}^{-1}$ and verify that this choice does not affect our measurements.
To improve the convergence of the fitting, we performed a two-step process.
We perform a pre-continuum fit to roughly estimate the amplitude/slope of the continuum in the first step.
Here we use the same procedures as those for the rest-frame UV continuum, with the filters only including the optical continuum and excluding the emission line filters.

We then conducted \texttt{emcee} fitting with four (or five) free parameters by converting continuum+line spectra into the fluxes in each band.
The priors of the amplitude/slope of the continuum are set to $\pm5\sigma$, those estimated in the first step.
We assign the prior of the line amplitudes to a linear range of [0,1000].
The upper boundary of the line amplitude is set to cover the brightest lines in the sample, at the fixed FWHM of $\sim500\,{\rm km}\,{\rm s}^{-1}$.
When we split \oiiihb\ into [O\,{\sc iii}] and H$\beta$, we limit $L_{{\rm H}\alpha+{\rm [NII]}}/L_{{\rm H}\beta}>2.86$ in the prior, which is based on the case-B recombination with $T_e=10^4\,{\rm K}$ and $n_e=10^2\,{\rm cm}^{-2}$ \citep{2006agna.book.....O}.
Figure \ref{fig:ex_fit} shows examples of the fitting results.
Thanks to the full NIRCam band coverage, our fitting captures the UV continuum, \hanii\ and \oiiihb\ emission lines, and the underlying optical continuum well for both significant ($\text{\ewhanii}\sim850\,$\AA) and insignificant ($\text{\ewhanii}\sim50$\AA) line emitters.
Effectively, our photometric measurement is sensitive down to ${\rm EW}\sim100\text{\AA}$, and therefore sources with $\lesssim100\text{\AA}$ tend to be treated as upper limits ($\sim10\%$ of the sample).

As our photometric line measurements are based on the $z_{\rm phot}$, we also account for its uncertainties.
We randomly pick 25 $z_{\rm phot}$ values from $p(z)$ derived by \texttt{Prospector}-$\beta$, and repeat the fitting procedures above.
We stack the posterior distribution from each individual fit to obtain the full posterior for each galaxy.
If the posterior distribution of the line amplitude continues increasing toward zero, we use 84th percentile values as upper limits.

In each fitted model spectrum from the stacked posterior distribution, we integrate a Gaussian function representing \hanii\ and \oiiihb\ (or two Gaussians when we split \oiiihb\ into ${\rm [OIII]}$ and ${\rm H}\beta$) and compute the line fluxes (${\rm erg}\,{\rm s}^{-1}\,{\rm cm}^{-2}$).
We estimate the underlying continuum at the line wavelength (${\rm erg}\,{\rm s}^{-1}\,{\rm cm}^{-2}\,{\text{\AA}}^{-1}$) from the best-fit power-law continuum, and calculate the observed-frame line equivalent widths (${\rm EW}_{\rm obs}$) by dividing line fluxes by the continuum at the line wavelength.
We then convert it to the rest-frame EW using ${\rm EW}_{\rm rest}={\rm EW}_{\rm obs}/(1+z)$.
From the UV luminosity derived from the redshift and $m_{\rm UV}$ at $\lambda_{\rm rest}=1500$\AA, we also calculate the luminosity ratio between emission lines and rest-frame 1500\AA\ UV continuum ($L_{{\rm H}\alpha+[{\rm N\,II}]}/\nu L_{\rm 1500}$ and $L_{{\rm [OIII]}+{\rm H}\beta}/\nu L_{\rm 1500}$, hereafter \rha\ and \roiii, respectively, which we collectively refer to as \rline).

We correct our \rline\ measurements for dust attenuation using the best-fit \texttt{Prospector}-$\beta$ model for each galaxy.
In these fits, dust is characterized by the two-component model in \citet{2000ApJ...539..718C} with a flexible dust attenuation curve \citep{2009A&A...507.1793N}, and described in three parameters, power-law index for a \citet{2000ApJ...533..682C} extinction curve ($n$), optical depth of diffuse dust ($\hat{\tau}_{\rm dust,2}$), and ratio between the optical depths of birth cloud dust and diffuse dust ($f_{\hat{\tau}}=\hat{\tau}_{\rm dust,1}/\hat{\tau}_{\rm dust,2}$, see \citet{2024ApJS..270...12W} for more detail).
The correction factor in \logrha\ (\logroiii) is calculated by taking the ratio between the attenuation at $\sim1500\text{\AA}$ and at $\sim6563\text{\AA}$ ($\sim5007\text{\AA}$).
The median and $1\sigma$ uncertainties of the correction factors of \logrha\ applied for individual galaxies are $-0.17_{-0.28}^{+0.10}$, $-0.12_{-0.17}^{+0.08}$, and $-0.11_{-0.16}^{+0.07}$ for the $z\sim3$, 5, and 6 samples, respectively.
For \logroiii, the correction factors are $-0.14_{-0.27}^{+0.09}$, $-0.10_{-0.15}^{+0.07}$, $-0.09_{-0.15}^{+0.06}$, and $-0.09_{-0.32}^{+0.06}$ for the $z\sim3$, 5, 6, and 8 samples, respectively.
At the stellar mass range focused on in this paper, dust extinction is not significant \citep[$A_V\lesssim0.5$, see also, ][]{2023ApJ...954..157S}, and the correction does not significantly affect our conclusions (see Appendix \ref{appendix:dust_corr} for further tests).

\subsection{Toy models of the star formation histories}\label{subsec:model}

To compare our measurement with SFHs on the population level, we create toy models using \texttt{fsps} \citep{2010ascl.soft10043C}.
The concept of the population-level constraint is demonstrated in \citet[][see also, \citealt{2025arXiv250616510M}]{2025arXiv250415255W}.
This approach attempts to model the distributions of parameters tracing different time scales in a given complete sample of galaxies as opposed to directly modeling individual galaxies. 
This approach assumes all of the galaxies in the sample have similar SFHs.
In \citet{2025arXiv250415255W}, they use three independent quantities to constrain the population-level SFH, ${\rm H}\alpha$ luminosity, UV luminosity, and Balmer break strength, since they are sensitive to the different time-scales of star formation ($\sim10\,{\rm Myr}$, $\sim100\,{\rm Myr}$, and $>100\,{\rm Myr}$, respectively).
Because measuring the Balmer break strength accurately only from the photometry is difficult due to the contamination of emission lines at $\sim4200$\,\AA, we focus on the luminosity ratio (\rha\ and \roiii) and ${\rm EW}_{\rm rest}$ (hereafter \ewha\ and \ewoiii) in this paper.

Here we briefly summarize the \texttt{fsps} model construction, which is similar to that in \citet{2025arXiv250616510M}.
We refer to \citet{2025arXiv250616510M} for more details on the \texttt{fsps} models.
We use the MILES stellar library \citep{2006MNRAS.371..703S}, MIST isochrones \citep{2016ApJS..222....8D,2016ApJ...823..102C}, and a \citet{2003PASP..115..763C} IMF. 
Input SFHs are characterized by periodic bursts and are parameterized as the combination of common exponential and double-sided natural exponential increase+decay as follows, 
\begin{align}
    &\text{SFR}(t, \tau, A_{\rm burst}, \delta t, \alpha) =\\\nonumber
&\left\{
    \begin{array}{lr}
        10^{\alpha t}\ A_{\rm burst}\ e^{[t\ (\text{mod } \delta t) - \delta t/2]/\tau}& \text{if } t\ (\text{mod } \delta t)< \frac{\delta t}{2} \\
        10^{\alpha t}\ A_{\rm burst}\ e^{-[t\ (\text{mod } \delta t) - \delta t/2]/\tau}& \text{if } t\ (\text{mod } \delta t) > \frac{\delta t}{2} \\
\end{array}
\right. .
\end{align}
The models have four parameters: decay time $\tau$ and amplitude $A_{\rm burst}$ of each burst, time interval between burst cycles $\delta t$, and overall slope $\alpha$.
We fix the decay time to $\tau=\delta t/30$.
For the remaining three parameters, we choose $A_{\rm burst}=0.8$, 1.4, 2.0, 2.6, 3.2\,dex, $\delta t=25$, 100, 200, 400, 800\,Myr and $\alpha=-0.004$, $-0.001$, 0.000, 0.001, 0.004.
To sample all phases of the burst cycle within 200 models, we shift $1\,{\rm Myr}$ in $\delta t=25$, 100, 200\,Myr models, $2\,{\rm Myrs}$ in $\delta t=400\,{\rm Myr}$ models, and $4\,{\rm Myrs}$ in $\delta t=800\,{\rm Myr}$ models and input model SFHs into \texttt{fsps}.
We assume the star formation continues for 800\,Myr, which corresponds to the star formation starting at $z\sim20$ for $z\sim6$ galaxies.
For the model spectra obtained from \texttt{fsps}, we fit a power-law and a Gaussian to the \texttt{fsps} spectra and calculate EWs and \rline\ in the same manner as that used for PRISM spectra.
As \texttt{fsps} spectra have better spectral resolution than that of PRISM spectra, we assume triple Gaussians for [O\,{\sc iii}]\,$\lambda\lambda$4959,5007+H$\beta\,\lambda$4861 and H$\alpha\,\lambda6563$+[N\,{\sc ii}]\,$\lambda\lambda6548,6583$ and sum all lines.

As the \texttt{fsps} SFH model construction takes significant computational resources, we fix the stellar/gas-phase metallicities and ionization parameters to $Z=0.1\,Z_{\odot}$ and $\log U_{\rm ion}=-1.5$ for comparison with the \hanii\ measurements (\rha\ and \ewha) roughly based on the average metallicity in our target mass range of $\log M_{\ast}[M_{\odot}]=8.0$--9.5 at $z=3$--9 \citep[e.g.,][]{2023ApJ...955...54S,2023ApJS..269...33N,2023MNRAS.518..425C}.
We also fixed the ionizing photon escape fraction $f_{\rm esc}=0$, as recent studies suggested that $f_{\rm esc}$ is typically as small as $\sim0.1$ at $z\sim4$--9 \citep[e.g.,][]{2023A&A...672A.155M,2024A&A...685A...3M}.
For comparison, we also calculate SFH models without bursts by simply excluding $A_{\rm burst}e^t$ terms from Eq. (1).

The variations of the nebular conditions, such as the metallicity ($Z$) and ionizing parameters ($U_{\rm ion}$), add additional scatter to \rline\ and EW distributions.
We estimated the effect of these variations by inputting a variety of $Z$ and $U_{\rm ion}$ values with constant SFH into \texttt{fsps}.
We choose the metallicity of $Z=0.05$--$0.5\,Z_{\odot}$ and the ionization parameter of $\log U_{\rm ion}=-2.0$--$-1.3$ \citep[e.g.,][]{2023ApJ...955...54S,2023ApJ...952..167R}.
We find \logrha\ and \ewha\ varies $\sim0.1$, $\sim200\,$\AA, $\sim0.2$ and $\sim100\,$\AA, respectively.
We virtually reproduced the scatter due to the variation of the metallicity and ionization by simply convolving a 1D Gaussian function with $\sigma_{\text{\logrha}}=0.05$, $\sigma_{{\rm EW},{{\rm H}\alpha}}=50\,$\AA\ into our corresponding \texttt{fsps} model distributions.
We will present the demonstration of our \texttt{fsps} models in Figure \ref{fig:modeldemo} (see also Figure \ref{fig:badmodelcomp} in Appendix \ref{appendix:noburst_models} for the effect of the $Z$ and $U_{\rm ion}$ variations).

%
%
%
%
%
%
\section{Results}\label{sec:results1}

\subsection{Robustness of the photometric measurements}\label{subsec:robustness}

To quantify the accuracy and precision of our photometric line measurements, we compare the line and continuum luminosities with those from NIRSpec/MSA spectra.
For the comparison spectroscopic measurements, we utilize the DR4 UNCOVER spectra \citep{2025ApJ...982...51P} with default drizzling.
We first limit H$\alpha$ and \oiii\ comparison sample with $z_{\rm spec}=2.7$--$7.0$ and \oiii\ comparison sample with $z_{\rm spec}=7.0$--$9.6$, given the PRISM's wavelength coverage, resulting in $\sim400$ galaxies.
As some spectra exhibit only tentative detection of the continuum, we use robust sources with $>10\sigma$ continuum detection for this comparison.
Most of the sources ($\sim90\%$) that do not satisfy this continuum criterion are very low mass ($\log M_{\ast}[M_{\odot}]<8.0$) or significantly attenuated by dust ($A_V>1.0\,{\rm mag}$).
The final comparison sample includes $\sim250$ galaxies.
We then perform a power-law and a Gaussian profile fitting in similar ways to those for the photometric measurements.
We applied an absolute flux correction to the MSA spectra by modeling the difference between the spectra and available NIRCam bands as a 5th-order polynomial.
If the emission lines are not detected at $5\sigma$ in the closest channel to the rest-frame line wavelength $\times(1+z_{\rm spec})$, we calculate an upper limit with $5\sigma$ value for line amplitude and ${\rm FWHM}=200\,{\rm km}\,{\rm s}^{-1}$.

%
%
%
%
%
%
\begin{figure}[tb]
\begin{center}
\includegraphics[width=8.8cm,bb=0 0 1000 650, trim=0 1 0 0cm]{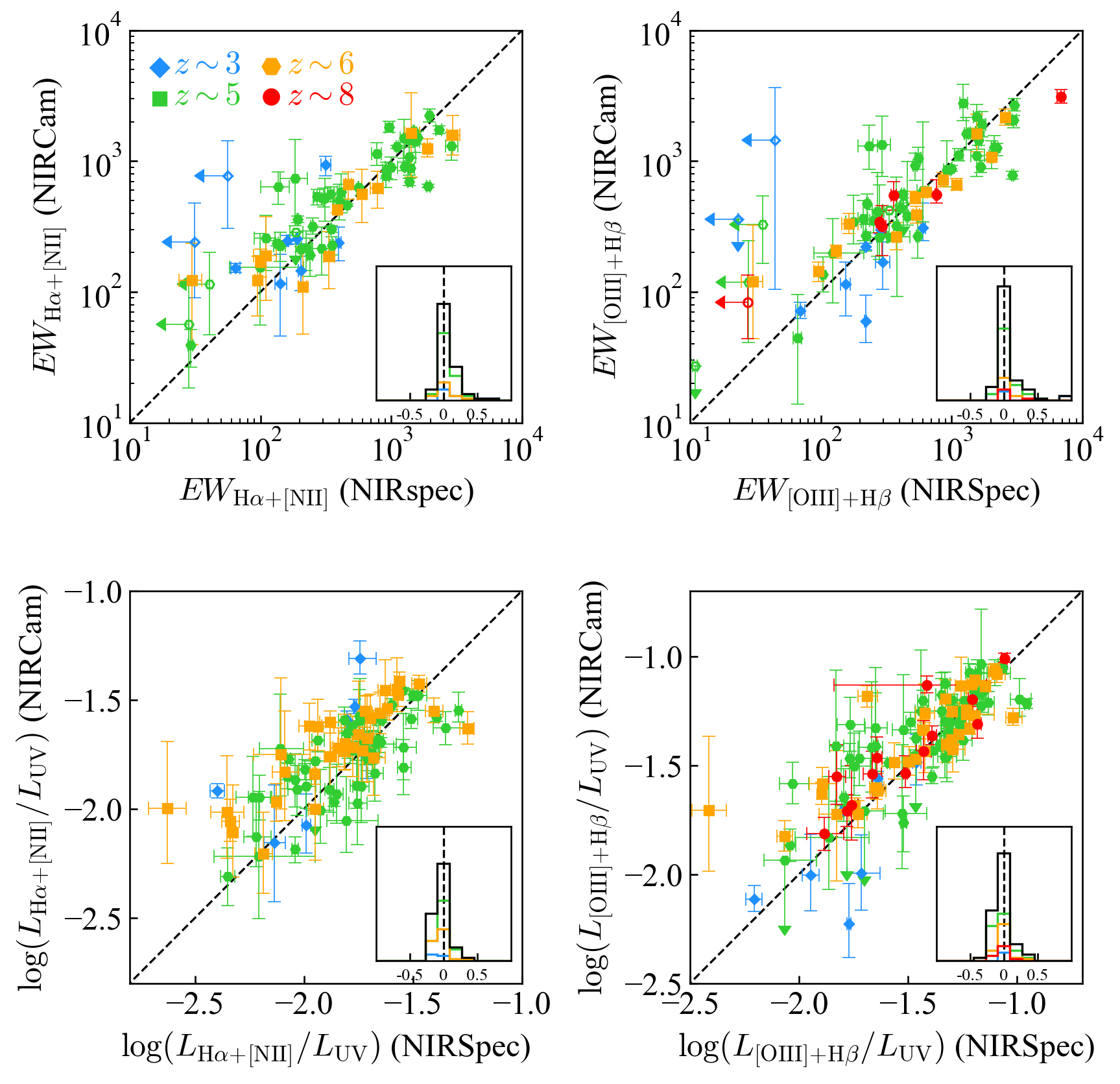}
\caption{Comparison between photometric and spectroscopic measurement of EWs (top) and \rline\ (bottom). \hanii\ and \oiiihb\ measurements are shown in the left and right panels, respectively. Different colors and markers correspond to the redshift samples (Section \ref{subsec:sample}). Inserted panels indicate the normalized deviation between NIRCam and NIRSpec measurements, as written by $\Delta=({\rm NIRCam}-{\rm NIRSpec})/{\rm NIRSpec}$. Our photometric measurements agree well with the spectroscopic measurements.}
\label{fig:specphoto}
\end{center}
\end{figure}
%
%
%
%
%
%

%
%
%
%
%
%
\begin{figure*}[tb]
\begin{center}
\includegraphics[width=18cm,bb=0 0 1000 650, trim=0 1 0 0cm]{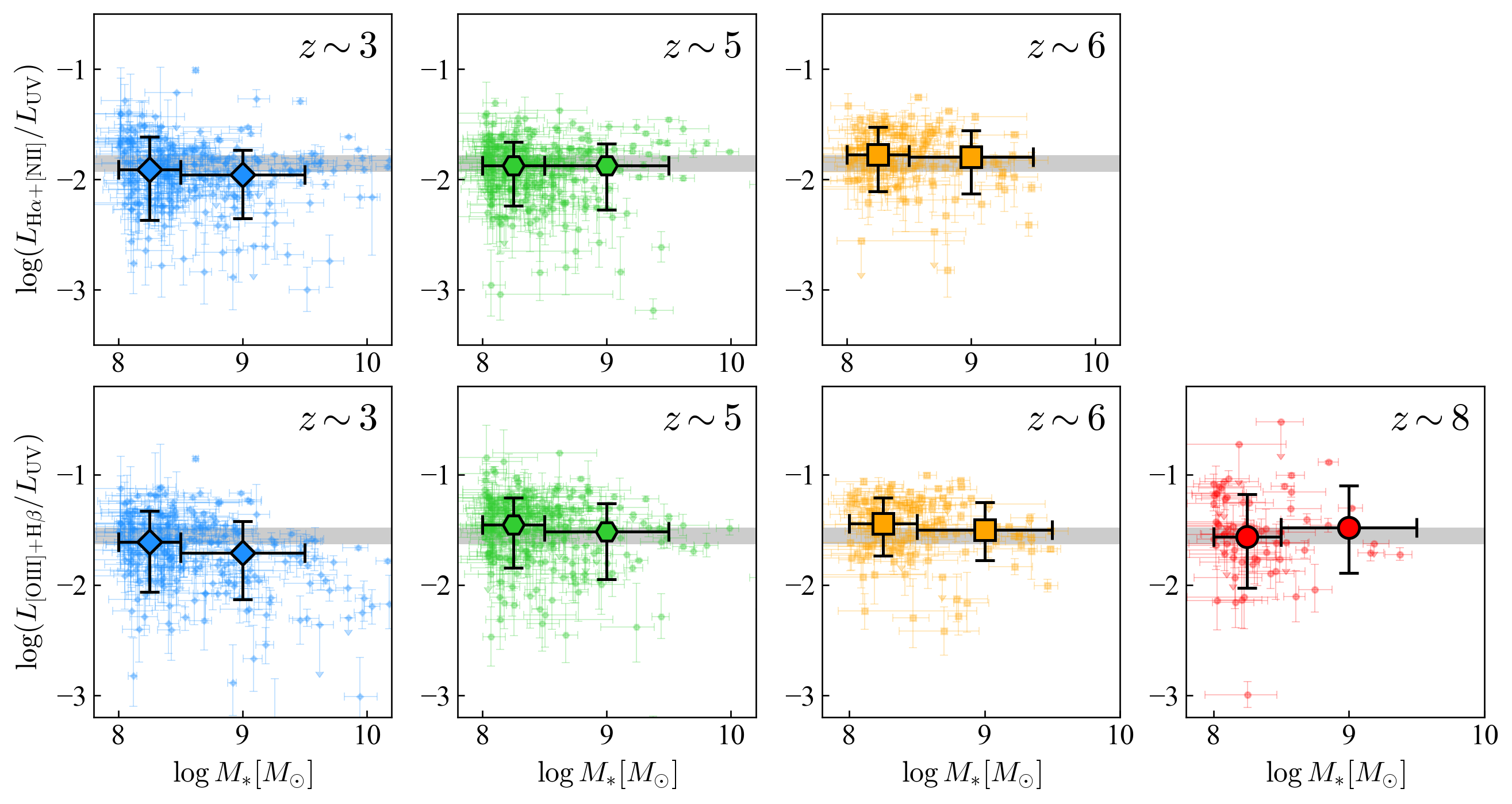}
\caption{\logrha\ (top panels) and \logroiii\ (bottom panels) as a function of stellar mass. From left to right, we show the $z\sim3$, 5, 6, and 8 subsamples. Small markers indicate individual galaxies, and large markers with black edges show the 16-50-84th percentiles of the stellar mass bins of $\log M_{\ast}[M_{\odot}]=8.0$--8.5 and 8.5--9.5. Gray shaded regions in the top panels and bottom panels show the range of \rha\ equilibilium value (1/85--1/60) and the range multiplied by two (see text), respectively. Here, we use the median value of the 16th-50th-84th percentile values among 500 posterior distributions for individual galaxies. There is no significant mass dependence of the scatter at the mass range of $\log M_{\ast}[M_{\odot}]>8.0$. \rha\ and \roiii\ exhibit similar trends specifically at $z>4$, suggesting usefulness of the \oiiihb\ lines as a tracer of the burstiness.}
\label{fig:masscomp}
\end{center}
\end{figure*}
%
%
%
%
%
%

%
%
%
%
%
%
\begin{figure*}[tb]
\begin{center}
\includegraphics[width=18cm,bb=0 0 1000 650, trim=0 1 0 0cm]{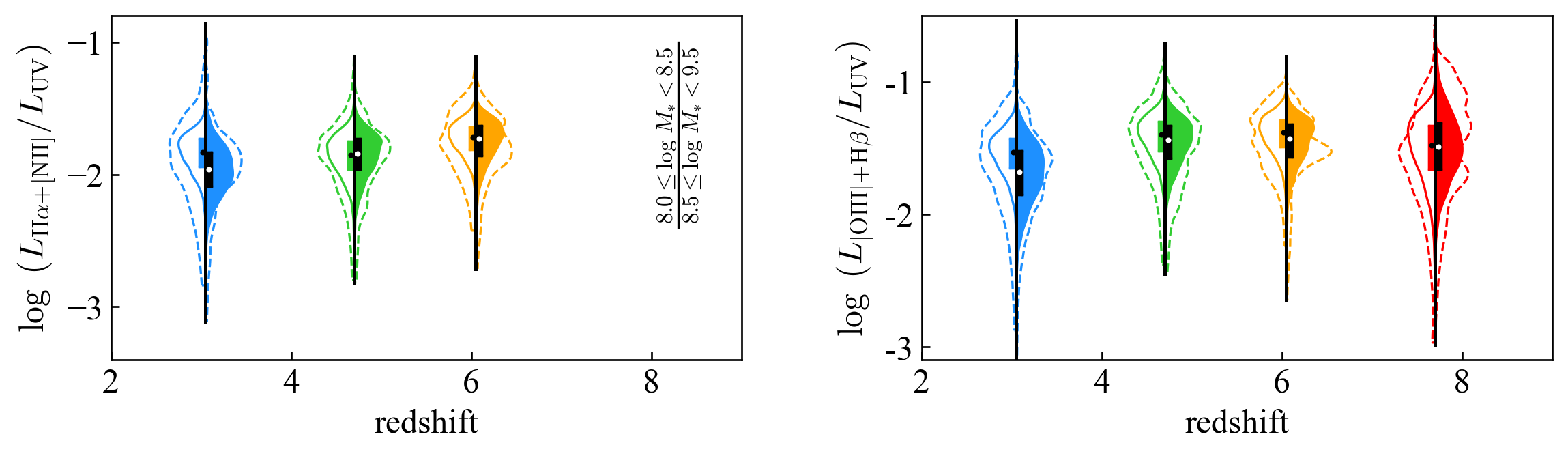}
\caption{Distribution of the \rha\ (left panel) and \roiii\ (right panel) of the mass-complete samples as a function of redshift. The left and right sides of the representative redshift exhibit the Bayesian hierarchical model of low-$M_{\ast}$, high-$M_{\ast}$, respectively. The dashed lines on each side correspond to the observed distribution used for the Bayesian modeling.}
\label{fig:zevolution1}
\end{center}
\end{figure*}

Figure \ref{fig:specphoto} shows comparisons between our photometric and spectroscopic measurements.
Overall, our photometric measurements reproduce the spectroscopic measurements well.
There is no significant systematic offset between photometric and spectroscopic measurements, as the normalized deviation $\Delta=({\rm NIRCam}-{\rm NIRSpec})/{\rm NIRSpec}$ of $\log$\ewha, $\log$\ewoiii, \logrha, and \logroiii\ are $\Delta \text{$\log$\ewha}=0.02_{-0.07}^{+0.11}$, $\Delta \text{$\log$\ewoiii}=0.01_{-0.07}^{+0.11}$, $\Delta \text{\logrha}=-0.05_{-0.08}^{+0.10}$, and $\Delta \text{\logroiii}=-0.03_{-0.12}^{+0.09}$, respectively.
We confirmed that the $\Delta$ value dependence on the stellar mass is not significant, mostly within the $1\sigma$ uncertainties derived in the full sample.
For the redshift dependence, we find the $z\sim3$ sample exhibits $\sim50\%$ larger scatter in the $\Delta$ value. 
This large scatter may be due to the filter coverage, slit-loss correction, or the low line luminosity. 
As medium-bands are distributed more densely in the LW than SW, empirical measurements have better sensitivity on \hanii\ and \oiiihb\ at $z\sim4$--7.
The slit-loss correction is another potential cause of the large scatter, as lower-$z$ galaxies tend to have larger sizes \citep[e.g.,][]{2014ApJ...788...28V,2023ApJ...951...72O} and show clear color gradients within single galaxies \citep[e.g.,][]{2019ApJ...877..103S}.
The line luminosity, specifically \oiiihb, in high-mass galaxies ($\log M_{\ast}[M_{\odot}]\gtrsim9$) gets faint due to the metallicity \citep[$Z\gtrsim0.2Z_{\odot}$][]{2023ApJS..269...33N,2023MNRAS.518..425C}.
These observational factors may lead to a large scatter in the $z\sim3$ sample (see Section \ref{subsec:zevo}).

\subsection{Observed \rline\ variations and trends}\label{subsec:rline}
We show \logrha\ and \logroiii\ as a function of the stellar mass in Figure \ref{fig:masscomp}.
\logrha\ mostly ranges from $\sim-1.5$ to $\sim-2.5$, and 50th percentile values ($P^{50}_{ \text{\logrha}}$) are consistent with the equilibrium value assuming constant SFH for $>100\,{\rm Myr}$, $-1.93\leq\text{\logrha}\leq-1.78$ \citep{2023ApJ...952..133M}.
The variation of \logrha, characterized by the difference between the 84th percentile and the 16th percentile (hereafter \ppha), is typically $\sim0.6$ (see Figure \ref{fig:zevolution2}).
Here we calculate \ppha\ in 1000 realizations from the posterior of the individual line fitting.
There is no significant dependence of $P^{50}_{\text{\logrha}}$ and \ppha\ on the stellar mass at any redshift samples.

The bottom panels of the Figure \ref{fig:masscomp} illustrate \logroiii\ instead of \logrha. 
Their 50th percentile values ($P^{50}_{\text{\logroiii}}$) mostly fall within a range scaled by a factor of 2 from the equilibrium value of \rha, as shown in gray shaded regions.
This can be simply explained by ${\rm H}\alpha/{\rm H}\beta\sim3$ and typical [O\,{\sc iii}]/H$\beta$ ($\sim6$) at $z>3$ \citep{2023ApJ...955...54S,2023ApJS..269...33N,2023MNRAS.518..425C}.
The variation of \logroiii, \ppoiii, is very similar to \ppha\ in redshift samples at $z\sim3$, 5, and 6.
As in the case of \rha, there is no significant dependence of $P^{50}_{\text{\logroiii}}$ and \ppoiii\ on the stellar mass at any redshift.
The $z\sim8$ sample may show larger \ppoiii\ than that at $z\sim3$, 5, and 6.

These trends are preserved in a first order if we plot \logroiii\ and \logrha\ as a function of the apparent magnitude in the optical wavelength ($m_{\rm opt}$) instead of the stellar mass, while there are several minor differences potentially due to the sample selection, like a enhancement of $P^{50}$ from a inclusion of low-mass bursty galaxies (see Appendix \ref{appendix:fluxcomp}).
This consistency implies that our results are not primarily driven by the prior stellar mass estimates derived using \texttt{Prospector}-$\beta$.

\subsection{Modeling the intrinsic scatter of \rline\ and its dependence on mass and redshift}\label{subsec:zevo}

In Section \ref{subsec:rline}, we found that there is no clear mass or redshift dependence of \rline.
Here, we use a Bayesian modeling framework which explicitly takes measurement errors into account, allowing us to derive the intrinsic distribution of \rline\ as shown in \citet{2020ApJ...893..111L} \citep[see also,][]{2025arXiv250406334P}.
Here, we assume that the intrinsic distribution of \rline\ is represented by an asymmetric Gaussian as follows,
\begin{align}
    &p(x) =\\\nonumber
&\left\{
    \begin{array}{lr}
        \dfrac{\sqrt{2}}{\sqrt{\pi}(\sigma^{-}_{\rm bin}+\sigma^{+}_{\rm bin})}\exp\left[-\dfrac{1}{2}\left(\dfrac{x-\mu_{\rm bin}}{\sigma^{-}_{\rm bin}}\right)^2\right] & (x < \mu_{\rm bin}) \\
        \vphantom{\rule{3pt}{5ex}}
        \dfrac{\sqrt{2}}{\sqrt{\pi}(\sigma^{-}_{\rm bin}+\sigma^{+}_{\rm bin})}\exp\left[-\dfrac{1}{2}\left(\dfrac{x-\mu_{\rm bin}}{\sigma^{+}_{\rm bin}}\right)^2\right] & (x > \mu_{\rm bin})
\end{array}
\right. .
\end{align}

where $\mu_{\rm bin}$, $\sigma^{+}_{\rm bin}$, and $\sigma^{-}_{\rm bin}$ are the mean, upper standard deviation, and lower standard deviation, respectively.
The variable $x$ represents physical parameters (\rha\ or \roiii) of the $N$ galaxies included in the sample.
Here, we separate the sample at each redshift into two different mass bins as in Figure \ref{fig:masscomp} at $\log M_{\ast}[M_{\odot}]=8.5$.
We take into account uncertainties of the individual line/continuum fitting, and then the likelihood is calculated as follows;
\begin{align}
\ln(P)\simeq\sum_{i=1}^{N}\ln\left(\dfrac{\Sigma_{j=1}^{M}p(x_{i,j})}{M}\right)
\end{align}
where $M$ is the index of the randomly sampled MCMC posteriors of the individual fitting; we incorporate 500 posteriors, i.e., $M=500$.
By accounting for the posteriors, we take into account the uncertainties of the individual line/continuum fitting.
Then we use \texttt{emcee} for the Bayesian modeling, with a flat prior in a linear range of $\mu_{\rm bin}=[-3,-1]$, $\sigma^{\pm}_{\rm bin}=[0.01,1.0]$.
We have confirmed that the lower/upper boundary of the prior does not affect the results, since the resulting scatter ($\sigma^{\pm}_{\rm bin}\sim0.4$--$0.8$) is sufficiently larger than the measurement uncertainties of the individual fitting ($\sim0.1\,{\rm dex}$) and smaller than unity.

Figure \ref{fig:zevolution1} shows the resulting \rha\ and \roiii\ distributions as a function of redshift.
To first order, the distributions are similar across all redshifts.
We find a small systematic offset towards low \rline\ for high-mass galaxies at $z\sim3$.
This may be due to systematically higher metallicities at $\log M_{\ast}[M_{\odot}]\sim9$ and $z\sim3$.
At this mass and redshift range, the gas-phase metallicity becomes $Z>0.4\,Z_{\odot}$, and both [O\,{\sc iii}] and H$\alpha$ get fainter than those at $Z\sim0.2\,Z_{\odot}$ \citep{2021ApJ...914...19S,2023ApJS..269...33N}.

\begin{figure*}[tb]
\begin{center}
\includegraphics[width=18cm,bb=0 0 1000 650, trim=0 1 0 0cm]{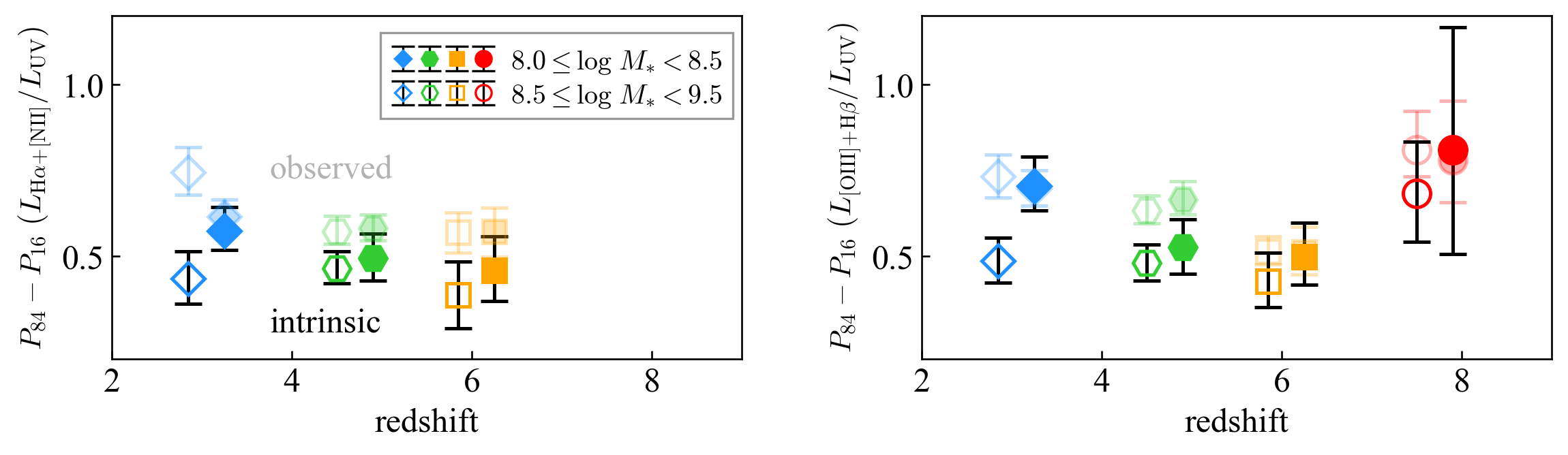}
\caption{Scatter in \rha\ (left panel) and \roiii\ (right panel) as a function of redshift. The open markers and filled markers show low-$M_{\ast}$ and high-$M_{\ast}$ samples, respectively. We introduce a small horizontal offset among the three samples to improve the visibility. Along with the Bayesian modeled intrinsic value, we also show the observed values at the semi-transparent markers.}
\label{fig:zevolution2}
\end{center}
\end{figure*}

Figure \ref{fig:zevolution2} illustrates the scatter in \rha\ and \roiii\ as a function of redshift. 
We show both the intrinsic \ppha\ and \ppoiii\ derived from the Bayesian model (black outlines) as well as the observed scatter (transparent), including measurement errors; filled markers show high-mass galaxies and open markers show low-mass galaxies.
In general, the intrinsic scatter shows no strong mass dependence.
The high-mass galaxies may have slightly larger intrinsic scatters, but they are consistent within 1$\sigma$. 
The exception is the $z\sim3$ sample, especially in \oiiihb, where high-mass galaxies have significantly more scatter.
As discussed in Section \ref{subsec:rline}, there is no significant dependence of the observed scatter on mass; however, the intrinsic scatter is potentially slightly larger for massive galaxies instead, while they are almost within $1\sigma$ uncertainties.
The reason for this potential positive mass dependence is not clear, but it may be due to the insufficient dust correction, as (1) the dependence gets weaker if we exclude the objects with relatively large dust extinction correction (Appendix \ref{appendix:dust_corr}), and (2) a larger dust amount is generally expected for higher-mass galaxies \citep[e.g.,][]{2020ApJ...902..112B,2020A&A...643A...4F,2024ApJ...971..161M}.

There is no significant evolution at $z<7$ for either \hanii\ or \oiiihb, which is consistent with the recent spectroscopic measurement \citep{2025ApJ...994...14P}, though the $z\sim3$ high-mass sample exhibits slightly larger scatter than $z\sim5$ or $z\sim6$.
This large scatter is potentially due to the metallicity and/or dust correction as discussed above.
In addition to the similarities between \roiii\ and \rha\ distributions, the \ppoiii\ shares a similar redshift trend with the \ppha\ at $z\sim3$--7, suggesting \oiiihb\ is also a statistically useful tracer to derive \rline\ distribution at this redshift range.
This cross-check allows us to compare \rline\ distribution and scatter between $z\sim3$--7 and $z\sim8$.
The $z\sim8$ sample has $\sim1.5$ times larger \ppoiii\ than the lower redshift samples in $1\sigma$ confidence, independent of the mass bins, although the associated uncertainty is also larger.
We observe similar trends in the flux-complete sample (Appendix \ref{appendix:fluxcomp}). 
We will discuss the possible origin of this largest scatter at $z\sim8$ by combining the population model in the following Section \ref{subsec:modelcomp}.

%
%
%
%
%
%

\subsection{Comparing our observations with toy models in FSPS}\label{subsec:modelcomp}

%
%
%
%
%
%
\begin{figure*}[htbp]
\begin{center}
\epsscale{1.15}
\includegraphics[width=16cm,bb=0 0 1000 650, trim=0 1 0 0cm]{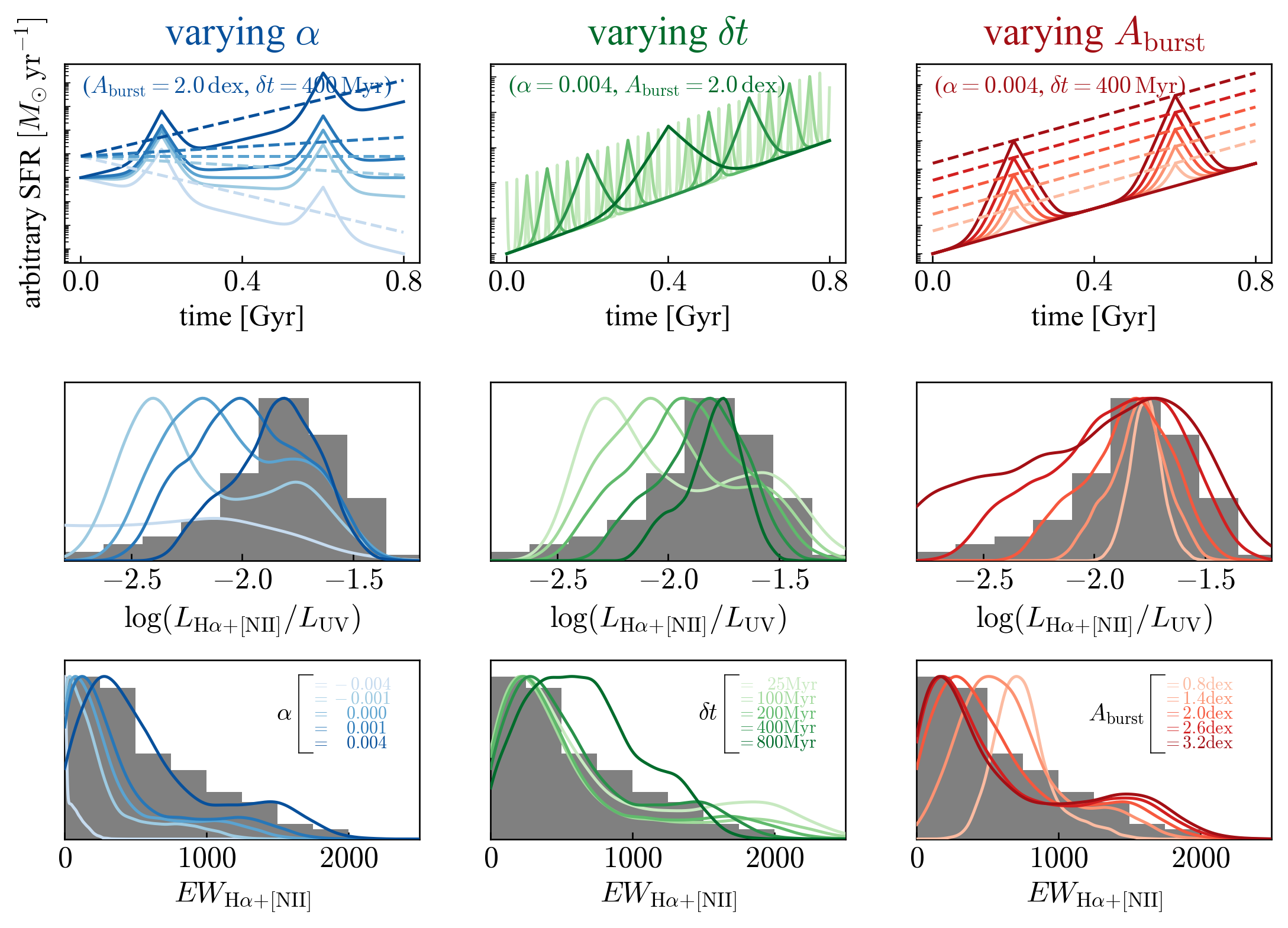}
\caption{Demonstration of how our population models change depending on the free parameters, $\alpha$, $\delta t$, and $A_{\rm burst}$. (top) SFHs as a function of the time since the Big Bang in Gyr. The model \logrha\ (middle) and \ewha\ (bottom) distributions, based on the SFHs indicated by the corresponding colors, are normalized to the maximum value for display purposes. The histograms in the middle and bottom panels indicate measured distributions of the $z\sim6$ sample with $\log M_{\ast}[M_{\odot}]=8.5$--9.5, where the distribution is inferred by stacking of posteriors in the individual fitting. Three parameters change the \logrha\ and \ewha\ distributions of the population, and therefore, we can constrain three free parameters by comparing overall distributions from observations and models.}
\label{fig:modeldemo}
\end{center}
\end{figure*}

In Section \ref{sec:results1}, we investigated the redshift and mass dependence of the \rline\ distributions, and found they are similar across our target stellar mass and redshift range.
Because \rline\ is sensitive to the SFH, this implies that the typical shape of the SFH in galaxies at $\log M_{\ast}[M_{\odot}]=8.0$--9.5 may be constant over $z=3$--9.
In this section, we explore families of SFH models that are consistent with our \rline\ measurements through comparison between observed \rline\ distributions and the toy models produced with \texttt{fsps} as described in Section \ref{subsec:model}.
Note again that, as in Section \ref{subsec:model}, we assume that galaxies included in the sample share similar SFHs when comparing the observed distribution of EWs and \rline\ with those of the toy models.
The changes of the \logrha\ and \ewha\ distributions owing to the three parameters of the toy model ($\alpha$, the slope of the SFH, $\delta t$, duration of the bursts, and $A_{\rm burst}$, the amplitude of the burst) are demonstrated in Figure \ref{fig:modeldemo}.
The overall trends are conserved if we use \oiiihb\ instead of \hanii, while \oiiihb\ allows slightly larger variations due to the metallicity (Appendix \ref{appendix:oiiihbcomp}).

\begin{figure*}[tb]
\begin{center}
\includegraphics[width=18cm,bb=0 0 1000 650, trim=0 1 0 0cm]{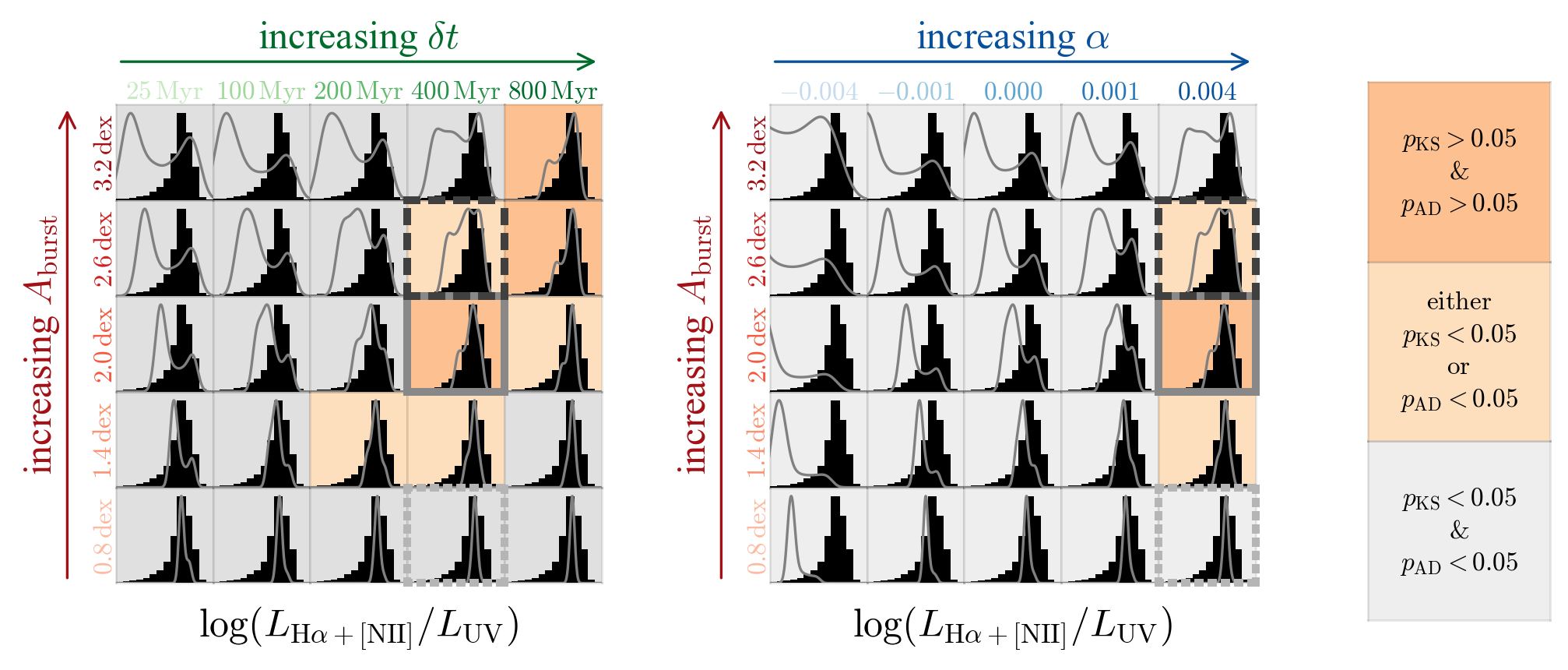}
\caption{\logrha\ distribution of the $z\sim6$ sample with $\log M_{\ast}[M_{\odot}]=8.5$--9.5. The range of the x-axis is the same as the middle row in Figure \ref{fig:modeldemo}. We overlay all \texttt{fsps} models with fixing $\alpha=0.004$ (left) or $\delta t=400\,{\rm Myr}$ (right) in gray curves. The \texttt{fsps} models change to increase $A_{\rm burst}$ from bottom to top for both left and right panels, increase $\delta t$ or $\alpha$ from left to right for left or right panels, respectively. Each panel is colored by the $p$-value of the KS and AD tests. As indicated in the three rightmost panels, $p_{\rm KS}>0.05$ and $p_{\rm AD}>0.05$, either $p_{\rm KS}<0.05$ or $p_{\rm AD}<0.05$, and both $p_{\rm KS}<0.05$ and $p_{\rm AD}<0.05$ are color-coded by thick orange, medium orange and gray, respectively. The observed distribution aligns well with the \texttt{fsps} model with long-duration, high-amplitude, and rising SFH. 
The panels indicating three comparison models used in Figure \ref{fig:zcompcomp} and \ref{fig:z12UVLF} are sorrounded in the corresponding gray colors.}
\label{fig:popfit}
\end{center}
\end{figure*}

In Figure \ref{fig:popfit}, we show the results of the Kolmogorov–Smirnov (KS) and Anderson–Darling (AD) tests between the observed and modeled \rline\ distributions for our representative mass and redshift sample ($z\sim6$, $\log M_{\ast}[M_{\odot}]=8.5$--9.5).
The KS test compares the cumulative distribution functions (CDFs) of two samples and is most sensitive to differences near the center of the distributions.
In contrast, the AD test also compares CDFs but places greater weight on differences in the tails, making it more sensitive to the edges of the distributions.
We performed these tests by randomly resampling the MCMC posterior of the individual objects 500 times to assess the robustness and variability, and evaluated the results based on the median values among the realization.
Three models have $p_{\rm AD}>0.05$ and $p_{\rm KS}>0.05$ in both \ewha\ and \logrha\ distributions (panels colored in thick orange), suggesting that the observed and modeled distributions are not statistically distinguishable and thus the models are plausible descriptions of the SFHs in this population. 
These three models are ($\alpha, \delta t, A_{\rm burst})=(0.004, 400\,{\rm Myr}, 2.0\,{\rm dex})$, ($0.004$, 800\,Myr, 2.6\,dex) and ($0.004$, 800\,Myr, 3.2\,dex).
Additional comparison in EW, and the limitations of the model are presented in Appendix \ref{appendix:comp_app}.

%
%
%
%
%
%

\begin{figure}[t]
\begin{center}
\includegraphics[width=7.0cm,bb=0 0 1000 650, trim=0 1 0 0cm]{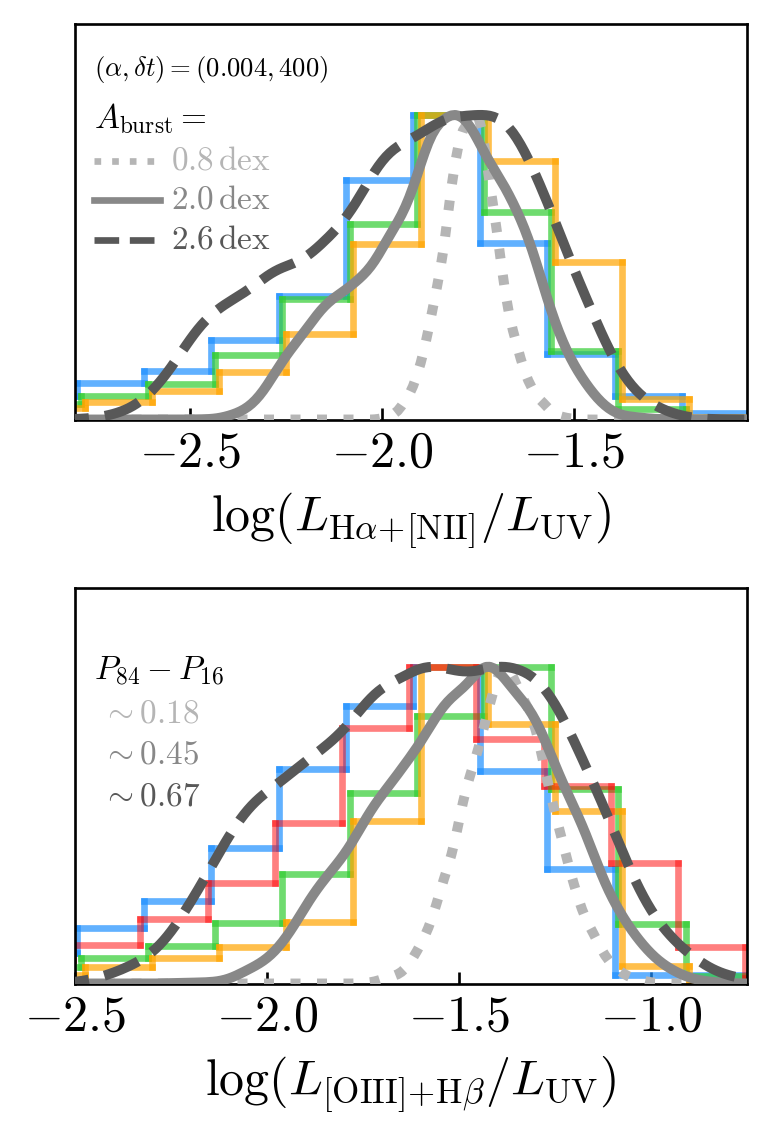}
\caption{Comparison of the observed \hanii\ and \oiiihb\ distributions from four redshift sabsamples ($z\sim3$, 5, 6, and 8 samples for blue, green, orange, and red, respectively) and \texttt{fsps} models in various $A_{\rm burst}=0.8$, 2.0, and 2.6\,dex with rising, long-duration ($\alpha$, $\delta t)=(0.004, 400$\,Myr) SFHs (gray colors).
$P^{84-16}$ values in different $A_{\rm burst}$ are also shown in the bottom panel.}
\label{fig:zcompcomp}
\end{center}
\end{figure}

Figure \ref{fig:zcompcomp} shows our \rline\ observations as well as three representative \texttt{fsps} toy models. 
Different colored histograms represent observations at different redshifts; consistent with Figure \ref{fig:zevolution1}, there is no significant redshift evolution. 
The grey lines show three toy models, each of which has a different burst amplitude ($A_{\rm burst}=0.8, 2.0, 2.6$ dex) but the same slope ($\alpha=0.004$) and timescale ($\delta t=400\,{\rm Myr}$). 
Here we focus on $A_{\rm burst}$, because the scatter in \rha\ and \roiii\ strongly depends on the amplitude of each burst ($A_{\rm burst}$) as shown in Figure \ref{fig:zcompcomp}.
Models without significant burstiness ($A_{\rm burst}=0.8\,{\rm dex}$) produce \rline\ distributions much narrower than our observations, while the burstiest models ($A_{\rm burst}=2.6\,{\rm dex}$) are typically too broad to describe our observations (other than \logroiiihb\ at $z\sim8$, the only bin with potential larger intrinsic scatter). 
For Section~\ref{sec:interp}, we will use this ``Goldilocks" $A_{\rm burst}=2.0\,{\rm dex}$ model that well-describes our $3<z<7$ observations to explore the effects of non-evolving burstiness on the $z>10$ UV luminosity function.

We explore whether the SFHs without bursts can reproduce the observed distribution in Appendix \ref{appendix:noburst_models}. 
Even with the reasonable variability in metallicity $Z$, ionization parameter $U_{\rm ion}$, and ionizing photon escape fraction $f_{\rm esc}$, the overall distribution in \texttt{fsps} model does not match to the observed distributions with $p$-values of $<0.001$, suggesting bursts are necessary \citep[see also,][]{2024MNRAS.535.1796B,2025A&A...697A..88L}.

%
%
%
%
%
%
%
%
%
%
%

%
%
%
%
%
%
\section{discussion}\label{sec:interp}

In Section \ref{subsec:zevo}, we showed that the intrinsic scatter in line strength to UV continuum does not depend strongly on stellar mass or redshift, indicating that SFHs may have similar shapes from $3<z<9$ and $8\lesssim\log M_{\ast}[M_{\odot}]\lesssim9.5$, and in Section \ref{subsec:modelcomp}, we showed that these observations are consistent with a population of galaxies that has rising SFHs with long-duration, high-amplitude bursts. 
In this section, we discuss the potential effect of these bursty SFHs on the abundance of UV-bright galaxies (Section \ref{subsec:z10}) and compare our results with previous studies (Section \ref{subsec:resultcomp}).
Throughout the Section, we assume the toy model with ($\alpha, \delta t, A_{\rm burst})=(0.004, 400\,{\rm Myr}, 2.0\,{\rm dex})$ unless otherwise specified, given that this model matches our observed \rline\ distributions.

\subsection{The abundance of UV-bright galaxies at $z>10$ in light of our SFH constraints at $z=3$--$9$}\label{subsec:z10}

\subsubsection{Enhancement of the $M_{\rm UV}$}

In this section, we explore the effects of (non-evolving) bursty SFHs on the abundance of UV-bright galaxies at $z>10$.
For long-duration bursts with $\delta t\gtrsim400\,{\rm Myr}$, galaxies at $z>10$ only have a single burst within the age of the Universe ($t_{\rm age}\lesssim500\,{\rm Myr}$).
As in Section \ref{subsec:model}, we created model SEDs in 200 phases of our bursty SFH in 2\,Myrs shift.
Here we fix these models to $z=11$, $Z=0.1\,Z_{\odot}$ and $\log U_{\rm ion}=-1.5$ and assumed that star formation continues $\sim340\,{\rm Myr}$, corresponding to star formation starting at $z=35$.
As an example, we normalized the SFHs to form a total mass of $\log M_{\ast}[M_{\odot}]=8.0$, which is likely to be typical stellar masses of the bright $z>10$ galaxies \citep[e.g.,][]{2025NatAs...9..155Z,2024Natur.633..318C,2024ApJ...972..143C}.
Then we computed magnitudes in JWST's broad bands.

\begin{figure}[tb]
\begin{center}
\includegraphics[width=8.6cm,bb=0 0 1000 650, trim=0 1 0 0cm]{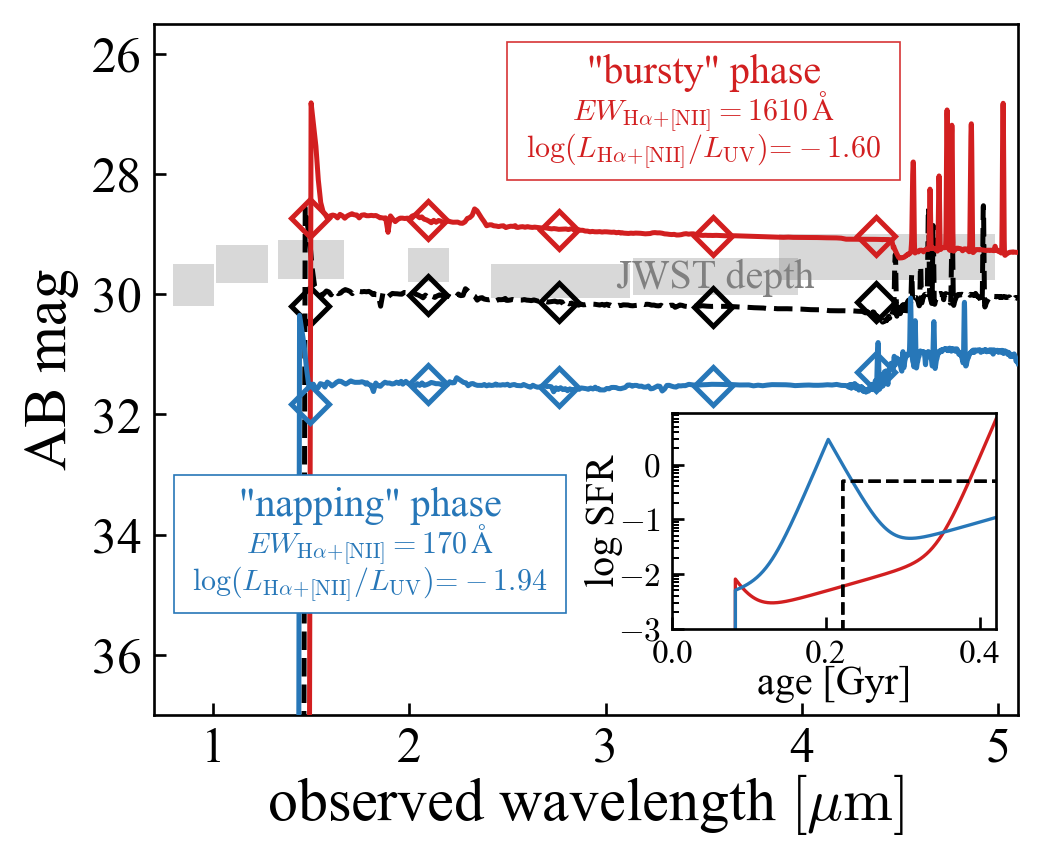}
\caption{The \texttt{fsps} modeled SEDs at $z=11$ in bursty (red) and napping (blue) phases, assuming SFHs derived in Section \ref{subsec:modelcomp}. 
Corresponding SFHs and NIRCam broadband magnitudes are shown in the inserted panels and the open diamonds. 
For comparison, the SED, NIRCam broadband magnitude, and SFH under 200\,Myr constant SFH are illustrated in black dashed lines. 
The survey depth on several surveys, such as UNCOVER/MegaScience \citep{2024ApJS..270....7W,2024ApJ...976..101S}, JADES, \citep{2023ApJS..269...16R}, and CEERS \citep{2025ApJ...983L...4F}, are displayed in gray shaded regions. We also show \logrha\ and \ewha\ at the bottom left of each panel. The $M/L$ in the rest-frame UV wavelength decreases, and the sources become detectable in the bursty phase at the given stellar mass.}
\label{fig:z10model}
\end{center}
\end{figure}

Figure \ref{fig:z10model} illustrates two representative SEDs at different phases of the SFH.
We also show a SED assuming 200\,Myr constant SFH as a reference (black).
In the phase close to the peak of the burst, the ``bursty'' phase (red), young O/B type stars are rapidly formed, and the mass-to-luminosity ratio ($M/L$) decreases. 
As a result, the brightness of the galaxy at UV wavelengths is boosted, and detectability is enhanced.
The decrease of $M_{\rm UV}$ reaches up to $\Delta M_{\rm UV}=-2\,{\rm mag}$ compared with a 200\,Myr constant SFH.
On the other hand, in the phase far from the burst, the ``napping'' phase (blue), slightly older A-type stars with higher $M/L$ dominate within the galaxy, making the galaxy significantly dimmer than in the ``bursty'' phase.

At fixed stellar mass ($\log M_{\ast}[M_{\odot}]\sim8.0$), we are only witnessing the ``bursty'' phase as predicted in theoretical simulations \citep[e.g.,][]{2023MNRAS.526.2665S,2024arXiv240910613N}.
In the typical survey depths with JWST (e.g., UNCOVER/MegaScience; \citealt{2024ApJS..270....7W,2024ApJ...976..101S}, JADES; \citealt{2023ApJS..269...16R}, and CEERS; \citealt{2025ApJ...983L...4F}), we found that galaxies with $\log M_{\ast}[M_{\odot}]\sim8.0$ are not detectable in F444W under a 200\,Myr constant SFH.
On the other hand, among 200 phases, 40\% are detectable (i.e., 40\% mass-completeness) thanks to the boost of $M_{\rm UV}$ from bursts.
We note that the bright $z\sim12$ galaxy from \citet[][]{2025NatAs...9..155Z} has $\text{\logrha}>-1.63$, consistent with the ``bursty'' phase and suggesting that currently confirmed $z>10$ galaxies may indeed be biased \citep[see also,][for the theoretical prediction]{2023MNRAS.526.2665S}.
Further direct observations of H$\alpha$ (or \oiiihb) emission line(s) will confirm the possible biases statistically.
Again, we note that this large effect on the observed UV magnitude occurs even assuming no increase in burstiness between $z\sim3$ and $z\sim11$.

%
%
%
%
%
%
%
%
%
%
%

%
%
%
%
%
%
\begin{figure*}[tb]
\begin{center}
\includegraphics[width=18cm,bb=0 0 1000 650, trim=0 1 0 0cm]{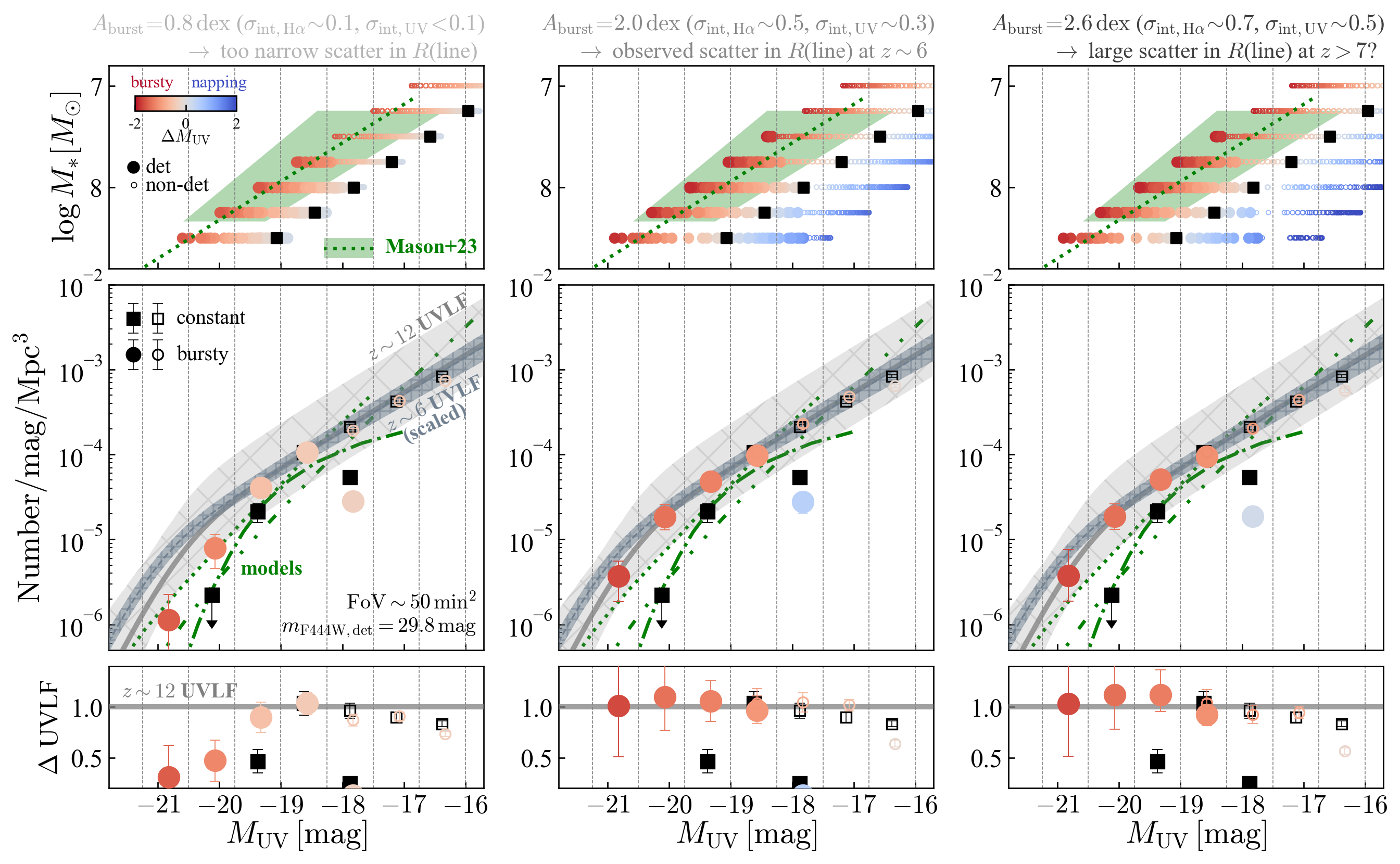}
\caption{The demonstration and resulting UV LF from our toy model based on the $A_{\rm burst}=0.8$ (left), 2.0 (middle), and 2.6\, dex (right). 
These amplitudes with $(\alpha, \delta t)=(0.004, 400\,{\rm Myr})$ correspond to $\sigma_{\rm int, UV}\sim0.01$, 0.3, and 0.5, and $\sigma_{\rm int, H\alpha}\sim0.1$, 0.5, and 0.7 (see Section \ref{subsec:resultcomp}).
[top row] The relationship between $M_{\rm UV}$ and $M_{\ast}$ in the bursty (circles) and constant SFH (squares). The circles from bursty SFHs are color-coded by $\Delta\,M_{\rm UV}$, defined as the difference between $M_{\rm UV}$ in a certain bursty SFH and $M_{\rm UV}$ in a 200\,Myr constant SFH. The circles are separated into large, filled markers and small, open markers based on their detectability in a typical JWST survey ($m_{\rm F444W}=29.8\,{\rm mag}$). The model from \citet{2023MNRAS.521..497M} is overlayed in green shades.
[middle row] The number densities, assuming bursty and constant SFH-based $M_{\rm UV}$ estimations as shown in the top panel, are illustrated in circles and squares, respectively. The filled markers indicate the number density of detectable galaxies with $m_{\rm F444W}<29.8\,{\rm mag}$, whereas open markers indicate the number density of all galaxies without consideration for detectability. 
At $M_{\rm UV}\lesssim-18.5\,{\rm mag}$, filled and open circles overlap, as all of these bright sources are detectable.
The number density from bursty SFH is color-coded by the average $\Delta M_{\rm UV}$ of the galaxies included in each $M_{\rm UV}$ bin.
The observed LF at $z\sim12$ \citep[light gray,][]{2025ApJ...980..138H}, scaled LF at $z\sim6$ \citep[dark gray,][]{2022ApJS..259...20H}, and several model UV LFs \citep[greens,][]{2015ApJ...813...21M,2020MNRAS.499.5702B,2023MNRAS.521..497M} are shown. 
[bottom row] The fractional difference between $z\sim12$ UV LF and our toy model UV LFs.
The model UV LF based on $A_{\rm burst}=2.0\,{\rm dex}$ reproduces the observed UV LF at $z\sim6$ as well as $z\sim12$, suggesting constant burstiness across $z\sim6$--12 is broadly consistent with the number density at the bright end.
While the potentially large $A_{\rm burst}$ at $z>7$ (Sec. \ref{subsec:modelcomp}) also gives a consistent UV LF, a small $A_{\rm burst}$ (e.g., 0.8\,dex) underestimates the number density at the bright end.}
\label{fig:z12UVLF}
\end{center}
\end{figure*}

\subsubsection{Effect on the UV LF}

This bias, where SFH burstiness boosts UV luminosity at fixed stellar mass, may explain the overabundance of UV-bright galaxies at $z>10$.
To demonstrate this effect, we create a model stellar mass function (SMF) at $z=12$ assuming a similar shape as at $z\sim9$ \citep[][]{2024MNRAS.533.1808W} and apply a linear scaling factor at the same amount as that between $z\sim6$ to $z\sim9$ \citep[i.e., assuming linear decrease of the stellar mass density across $z\sim6$-12,][]{2024MNRAS.533.1808W,2025ApJ...978...89H}.
Since the survey volume of JWST observations is limited, typically $\sim50\,{\rm arcmin}^2$ in deep or medium-deep surveys (correspnding to $\log \phi\,[{\rm Mpc}^{-3}\,{\rm dex}^{-1}]\sim-5.0$), high-mass galaxies with $\log M_{\ast}[M_{\odot}]>8.5$ are expected to be vanishingly rare.
We sample galaxies in the stellar mass range $\log M_{\ast}[M_{\odot}]=7$–8.5, weighted by the SMF in bins of width 0.25\,dex in $\log M_{\ast}$.
Next, we calculate $M_{\rm UV}$ assuming either our bursty SFH model or a 200\,Myr constant SFH at the given $M_{\ast}$. We compute $M_{\rm UV}$ for the three toy models shown in Figure~\ref{fig:zcompcomp}. 
For the bursty SFH case, we calculate the conversion factor from $M_{\ast}$ to $M_{\rm UV}$ at 200 phases in 2\,Myr steps, and randomly choose the phase on the SFH and convert $M_{\ast}$ into $M_{\rm UV}$. 
For the constant SFH case, we use uniform conversion factors from stellar mass to UV luminosity.
Finally, we constructed a luminosity function (LF) based on the number of galaxies at each $M_{\rm UV}$ bin.
Note that we adopt uniform metallicity of $Z=0.1\,Z_{\odot}$ (see Section \ref{subsec:model}) and do not apply any dust effect, which may be other important factors in understanding UV LF \citep[e.g.,][]{2023MNRAS.522.3986F,2024MNRAS.529.3563T}.

These procedures are illustrated in the top panels of Figure \ref{fig:z12UVLF}.
The $M_{\ast}$--$M_{\rm UV}$ relation used in the constant SFH is represented by the black squares, and those randomly picked from the bursty SFH are color-coded by the difference of $M_{\rm UV}$ from $M_{\rm UV}$ in the constant SFH case ($\Delta M_{\rm UV}$).
Bursty SFHs can cause $M_{\rm UV}$ to vary up to $\pm2$ magnitudes versus a constant SFH. 
Generally, in the bursty SFH case, galaxies in the ``bursty'' phase upscatter toward the bright $M_{\rm UV}$ bins, and those in the ``napping'' phase downscatter toward the faint $M_{\rm UV}$ bins. 
Filled circles represent galaxies that are detected in a typical deep$\sim$medium-deep extragalactic survey ($m_{\rm{F444W}}\le29.8$), and empty markers represent undetected galaxies. 
The burstiest phases are detectable even at low mass, while only the highest-mass napping galaxies are detected. 

The resulting toy model UV LF is shown in the middle panels of Figure \ref{fig:z12UVLF} in order of $A_{\rm burst}$ from left to right.
Each point is colored by the average $\Delta M_{\rm UV}$ for all galaxies in that $M_{\rm UV}$ bin. 
With a constant SFH, there are no objects brighter than $M_{\rm UV}\sim-20\,{\rm mag}$ -- these UV-bright objects would have to have much higher stellar masses than our mass cap of $\log M_{\ast}[M_{\odot}]=8.5$, and are extremely unlikely in a $\sim50\,{\rm arcmin}^2$ field of view. 
Note that our results do not change significantly if the mass cap is varied by $\sim0.5\,{\rm dex}$.
In the bursty SFH, only the most bursty galaxies ($\Delta M_{\rm UV}<-1.0$) occupy the bright end of the model UV LF ($M_{\rm UV}<-20\,{\rm mag}$). 
It is not until $M_{\rm UV}>-17\,{\rm mag}$ that the number of high-mass downscattered napping galaxies begins to balance the number of low-mass upscattered bursty galaxies, resulting in average $\Delta M_{\rm UV}\sim0.0$.
This picture also naturally explains an apparent dependence of the SFH on $M_{\rm UV}$ as presented in \citet{2024MNRAS.533.1111E}, since the fraction of the galaxies in ``bursty'' phase ($\Delta M_{\rm UV}<0$) get smaller at the fainter $M_{\rm UV}$ bin.

The shape of the bright end of the UVLF depends significantly on the assumed burstiness. 
The leftmost column in Figure \ref{fig:z12UVLF} shows the $A_{\rm burst}=0.8\,{\rm dex}$ toy model. 
While this model produced \rline\ distributions too narrow compared with our observations at $3<z<7$ (Figure \ref{fig:zcompcomp}), it matches the approximate range of $M_{\ast}$--$M_{\rm UV}$ used in the \citet{2023MNRAS.521..497M} models (top left in Figure \ref{fig:z12UVLF}). 
This model underpredicts the number of UV-bright galaxies, with offsets up to 0.5~dex at the bright end. In contrast, the toy models with $A_{\rm burst}=2.0\,{\rm dex}$ and $A_{\rm burst}=2.6\,{\rm dex}$ (middle and right panels in Figure \ref{fig:z12UVLF}) both reproduce the $z\sim12$ UVLF with no significant offsets. 
The $A_{\rm burst}=2.0\,{\rm dex}$ model reproduces the bulk of our observations at $3<z<7$, while the $A_{\rm burst}=2.6\,{\rm dex}$ model has better alignment for our $z\sim8$ observations. 
The inferred scatter in the UV absolute magnitude ($\sigma_{M_{\rm UV}}$) in the $A_{\rm burst}=2.0\,{\rm dex}$ and $A_{\rm burst}=2.6\,{\rm dex}$ models are $\sigma_{M_{\rm UV}}\sim1.1$ and $\sim1.3$, respectively, which have good agreements with the theoretical prediction at $z\sim10$--13 \citep{2024arXiv240504578K}, while they predict much higher $\sigma_{M_{\rm UV}}$ at $z\sim20$.
On the whole, these models suggest that no increase in burstiness from $z\sim3$ to $z\sim12$ is required to reproduce the observed large number of bright galaxies at high redshift: instead, several models of the high-redshift UVLF underestimate the amount of burstiness.
While in practice there may be additional factors, this simple toy model demonstrates that an SFH model consistent with our $z\sim3$--9 observations is already bursty enough to explain the large number of UV-bright galaxies at $z>10$.

%
%
%
%
%
%
\begin{figure*}[tb]
\begin{center}
\includegraphics[width=17cm,bb=0 0 1000 650, trim=0 1 0 0cm]{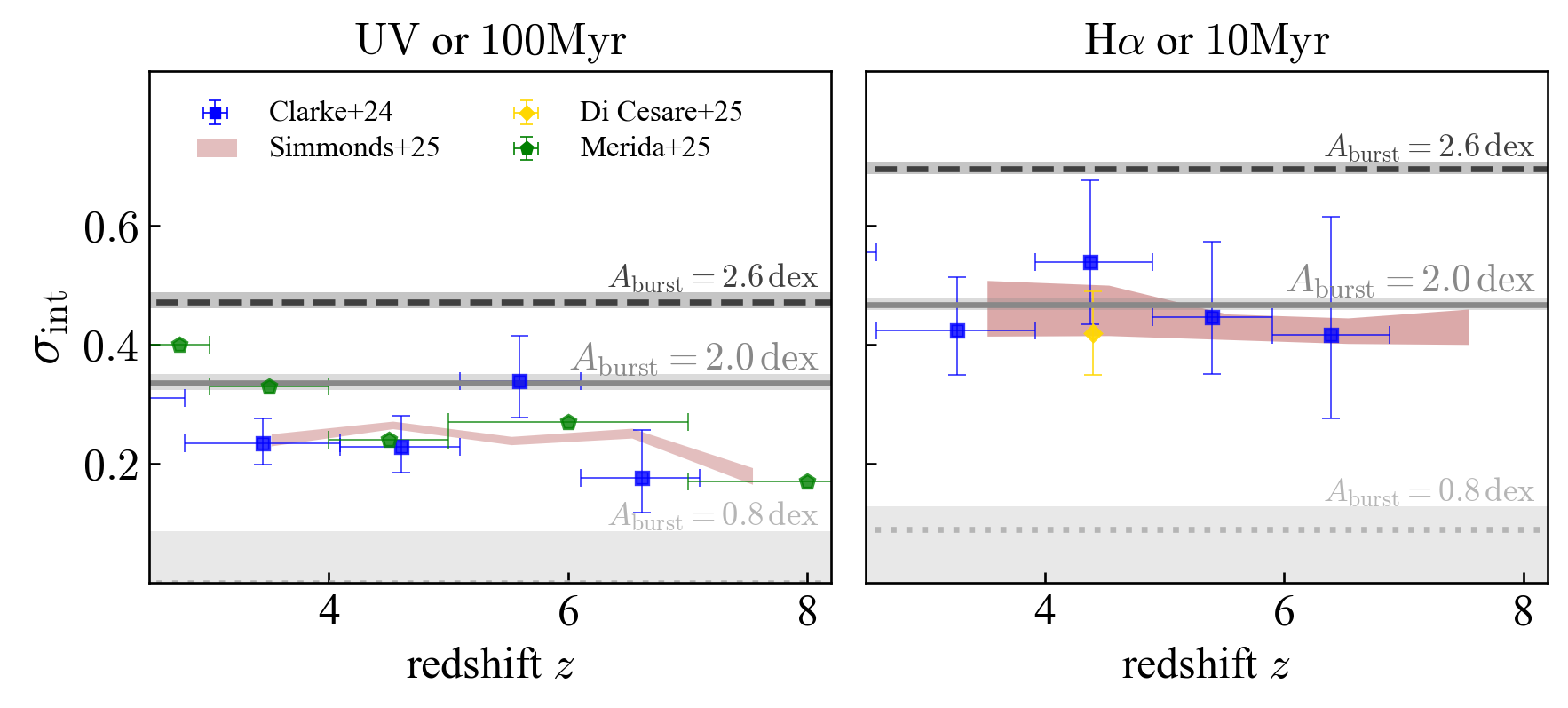}
\caption{Scatter of the SFMS based on the UV- ($\sigma_{\rm int, UV}$, left) and H$\alpha$-based SFR ($\sigma_{\rm int, H{\alpha}}$, right) as a function of the redshift. 
The scatters based on the $A_{\rm burst}=0.8$, 2.0, and 2.6\,dex are shown in gray colors corresponding to Figure \ref{fig:zcompcomp}, with several previous studies \citep{2024ApJ...977..133C,2025arXiv250804410S,2025arXiv250922871M,2025arXiv251019044D}.
The $\sigma_{\rm int, H{\alpha}}$ from the $A_{\rm burst}=2.0\,{\rm dex}$ model are broadly consistent with previous studies, while $\sigma_{\rm int, UV}$ is a bit larger than them, suggesting slightly smaller $A_{\rm burst}$ in their SFMS.}
\label{fig:MS_sigma}
\end{center}
\end{figure*}
%
%
%
%
%
%

%
%
%
%
%
%
\subsection{Comparison with previous studies and simulations}\label{subsec:resultcomp}

In Figure \ref{fig:masscomp} and \ref{fig:zevolution2}, we find no significant dependence of the observed $P^{50}_{\text{\logrha}}$ and \ppha\ on stellar mass at any redshift.
Generally, lower-mass galaxies are expected to be affected easily by the feedback from SNe, photoionization, and stellar winds because of their shallow potential well (i.e., they have efficient feedback). 
This efficient feedback would produce a large scatter in the intrinsic \rline\ distributions; however, these effects primarily dominate at lower masses than we focused on in this work ($\log M_{\ast}[M_{\odot}]\lesssim8$).
Indeed, the mass dependence of the feedback efficiency is not significant at our target mass range in some numerical simulations with realistic feedback \citep[e.g.,][]{2012MNRAS.421.3522H,2017MNRAS.466...88S}, and in some observational studies \citep[e.g.,][see also \citet{2015ApJ...809..147H} for an example of mass-dependent feedback efficiency]{2019ApJ...886...74M}. 
These results are consistent with our finding that the burstiness does not significantly depend on mass. 
In the local Universe \citep[e.g.,][]{2012ApJ...744...44W,2019ApJ...881...71E}, there is no significant mass dependence at $\log M_{\ast}[M_{\odot}]=8.0$--9.5, which is in line with results: the scatter only begins to increase significantly at $\log M_{\ast}[M_{\odot}]<8.0$.
Additionally, several previous observational \citep[e.g.,][]{2013ApJ...770...57B,2013MNRAS.428.3121M} and theoretical results \citep[e.g.,][]{2012MNRAS.421.3522H} suggest potential changes in feedback efficiency at either lower ($\log M_{\ast}[M_{\odot}]<8.0$) or higher mass ($\log M_{\ast}[M_{\odot}]>9.5$) regimes than we probe here.
At $\log M_{\ast}[M_{\odot}]<8.0$, short time, highly time-variable stochastic star formation than those at $\log M_{\ast}[M_{\odot}]=8.0$--9.5 is expected because stellar feedback from supernovae, radiation pressure, and stellar winds efficiently removes gas from the central regions \citep{2014MNRAS.445..581H}.
On the other hand, in sufficiently massive galaxies with $\log M_{\ast}[M_{\odot}]\gtrsim9.5$--10.0, deep gravitational potentials, stellar feedback cannot fully evacuate the gas, leading instead to a quasi-steady, self-regulated mode of star formation \citep[][]{2014MNRAS.445..581H,2018MNRAS.473.3717F}. 
Our focus mass range likely corresponds to a transitional regime between two distinct modes of star formation, and therefore, there may be no clear mass dependence \citep[e.g.,][]{2017MNRAS.466...88S}.
Covering a wider dynamic range in $M_{\ast}$ via further deep and wide observations is critical to constrain possible mass dependences of \rline.

Our toy modeling allows us to explore the types of SFHs that are most consistent with our observed \rline\ and EW distributions. 
As demonstrated in Figure \ref{fig:zcompcomp} and explored further in Appendix \ref{appendix:comp_app} and Figure \ref{fig:popfit}, our observations are most consistent with rising ($\alpha\geq0.001$), long-duration ($\delta t\geq200\,{\rm Myr}$), and high-amplitude ($A_{\rm burst}\geq2.0\,{\rm dex}$) SFHs. 
These results generally align well with \citet{2025arXiv250616510M}, where they measure Balmer break strength and H$\alpha$ EWs using NIRSpec/PRISM spectra of a flux-complete dwarf galaxy sample at $z\sim2$, then use similar population models to constrain their SFHs (see also, Section \ref{subsec:caveat}).
This consistency between $z\sim2$ and $3<z<7$ supports our finding that there is no clear redshift dependence of the population SFH over most of our target redshift range \citep[see also,][]{2025arXiv250804410S}. 
Naturally, we can reach the same conclusion in \citet{2025arXiv250616510M} about the physics driving these bursty SFHs: the variation of SF on $>100\,{\rm Myr}$ timescales can not be explained with the variation of giant molecular clouds (GMCs), but with gas cycling driven by SN feedback and cooling \citep[see also,][]{2020MNRAS.497..698T}. 
We find that the intrinsic scatter in \rline\ increases for our highest-redshift sample at $z\sim8$, which may imply that burstiness increases in the very early Universe, although these measurements are subject to uncertainties in metallicity.

Recently, several studies reported larger scatter in the SFMS when using short-timescale SFR tracers (H$\alpha$ or ${\rm SFR}_{\rm 10Myr}$ from non-parametric SED fitting) than when using long-timescale tracers such as rest-frame UV continuum or SED-fitting ${\rm SFR}_{\rm 100Myr}$ \citep[e.g.,][see also, \citealt{2012ApJ...754L..29W,2013ApJ...777L...8K,2015ApJ...815...98S,2015A&A...575A..74S}]{2024ApJ...977..133C,2025arXiv250804410S}.
These studies have typically interpreted this as evidence that short-timescale bursts are necessary -- apparently at odds with our finding that longer $>200\,{\rm Myr}$ bursts best describe our observations. 
However, the burst duration timescale ($\delta t$) is not directly comparable to the fluctuation timescale in these studies. 
The short-term fluctuation in these studies is almost equivalent to the necessity of bursts (expressed by the exponential increase+decay in equation 1) as suggested in Section \ref{subsec:modelcomp}, but is not equal to a small $\delta t$.

To compare our SFH models and previous studies about the SFMS more directly, we compute an intrinsic scatter of the SFMS based on the H$\alpha$ or UV-based SFRs ($\sigma_{\rm int, H{\alpha}}$ or $\sigma_{\rm int, UV}$).
Here, we utilize the $z\sim6$ sample as a representative and construct SFMS in a Monte-Carlo procedure.
First, we pick the $M_{\ast}$ of each galaxy from \texttt{Prospector} fitting, and scale the total mass produced by bursty SFHs into the $M_{\ast}$.
We choose a random phase for each object and compute $L_{\rm H\alpha}$ and $L_{\rm UV}$ using \texttt{fsps}.
We then convert $L_{\rm H\alpha}$ and $L_{\rm UV}$ into ${\rm SFR}_{\rm H\alpha}$ and ${\rm SFR}_{\rm UV}$ with \citet{1998ARA&A..36..189K} relation, and calculate $\sigma_{\rm int, H\alpha}$ and $\sigma_{\rm int, UV}$ in a same mannar with \citet[][]{2024ApJ...977..133C}.
As our calculation includes a random factor (i.e., phase of the bursty SFH), we repeat these procedures 1000 times and evaluate the probable $\sigma_{\rm int, H\alpha}$ and $\sigma_{\rm int, UV}$ values.
We computed $\sigma_{\rm int, H\alpha}$ and $\sigma_{\rm int, UV}$ based on the different $A_{\rm burst}$ with $(\alpha, \delta t)=0.004, 400\,{\rm Myr}$ as a comparison.

The comparison of $\sigma_{\rm int, H\alpha}$ and $\sigma_{\rm int, UV}$ with several previous studies is shown in Figure \ref{fig:MS_sigma}.
Our $A_{\rm burst}=2.0$~dex model produces 0.52 dex of scatter around the H$\alpha$ SFMS broadly consistent with the previous results \citep{2024ApJ...977..133C,2025arXiv250804410S,2025arXiv251019044D}.
This scatter around the UV-based SFMS (0.35~dex) is slightly higher than that of the other observations \citet{2024ApJ...977..133C,2025arXiv250804410S,2025arXiv250922871M}, suggesting slightly less bursty SFHs.
For instance, a model with $A_{\rm burst}=1.4\,{\rm dex}$ corresponds to $\sigma_{\rm int, UV}=0.23$, suggesting better consistency with the UV-based SFMS presented in the previous studies.
Overall, an ensemble of galaxies with rising SFHs and large variations above/below the main sequence can produce large differences in ${\rm SFR}_{\rm H\alpha}/{\rm SFR}_{\rm UV}$, even if the burst duration for any single galaxy is significantly longer than 100\,Myr. 
This in turn means that $\delta t$ is the most difficult parameter in our model SFH to constrain, even at a population level (see middle panel of Figure \ref{fig:modeldemo}).

\subsection{Potential caveats and future prospects}\label{subsec:caveat}

In this work, we showed that NIRCam medium-band measurements are one of the ideal datasets to construct a mass- (or flux-) complete sample in a moderate FoV ($\sim35\,{\rm min}^2$, Section \ref{subsec:sample}).
Our empirical line/continuum measurements are accurate with respect to spectroscopic measurements (Section \ref{subsec:robustness}).
Additionally, the photometric dataset enables us to avoid slit losses, which is an advantage over spectroscopic measurements.
Even with this high-quality data, there are still some potential caveats to our analysis.

\subsubsection{Observational caveats}\label{subsubsec:obscaveat}
First, it is challenging to accurately capture some spectral features from medium-band photometry alone, including measuring the Balmer break strength or faint emission lines (e.g., [O\,{\sc ii}]$\lambda$3727), and deblending some close metal emission lines (H$\alpha$+[N\,{\sc ii}] and [O\,{\sc iii}]+H$\beta$), as mentioned in Section \ref{subsec:photmeasure}. 

The Balmer break is sensitive to long-duration SFH, and may be important to improve our comparison between observations and models \citep[see][]{2025arXiv250616510M}.
The spectral line measurement of not only bright emission but also faint ones enables us to derive source-by-source variation of the nebular condition, such as $f_{\rm esc}$, $Z$, $U_{\rm ion}$, and dust attentuation \citep{2021ApJ...914...19S,2023ApJS..269...33N,2024MNRAS.529.3751C,2023ApJ...954..157S}.
We uniformly assume $f_{\rm esc}=0$ for modeling our population of bursty galaxies with \texttt{fsps}, while it does not significantly affect our results and conclusions.\footnote{Assuming $f_{\rm esc}\sim0.1\pm0.05$ (typical values at $z>4$), the \rline\ and EW get systemtically smaller by $\sim0.05\,{\rm dex}$ and $\sim10\%$, and broaden the \rline\ and EW distributions by $\sim0.025\,{\rm dex}$ and $\sim5\%$ (see top and bottom panels of Figure \ref{fig:badmodelcomp} in Appendix \ref{appendix:noburst_models} as an example of the $f_{\rm esc}$ effects).}
Both systematic and statistical variation in $Z$ and $U_{\rm ion}$ directly connect to that in \rline.
For instance, systematically high $U_{\rm ion}$ results in high \rline, which has a comparable effect with the slope of the SFHs ($\alpha$).
If the scatter in $U_{\rm ion}$ and $Z$ actually gets larger (or smaller) as a function of redshift, the scatter due to the SFH can be smaller (or larger), while this effect is expected to be minor even in a reasonably wide range of $U_{\rm ion}$ and $Z$ (see Section \ref{subsec:model}).
Additionally, more accurate measurements of the dust extinction would require Balmer decrements, although we corrected the dust extinction effect from the \texttt{Prospector} fitting and confirmed that the dust extinction correction does not significantly affect the modeling results (Appendix \ref{appendix:dust_corr}).

Deblending [O\,{\sc iii}] and H$\beta$ has the merit of enabling the detection of pure Balmer lines out to $z\sim9.5$.
[O\,{\sc iii}] lines have a stronger dependence than Balmer lines on power sources beyond star formation (e.g., AGNs and shocks) and on the nebular conditions (e.g., metallicity and ionizing radiation, see Appendix \ref{appendix:oiiihbcomp}).

Future large spectroscopic surveys using, for example, MSA/PRISM on JWST/NIRSpec toward a mass- (or flux-) complete sample will allow us to measure several spectral features which cannot be captured from photometry.

Second, a wider FoV and deeper observations are important to capture massive/bright and low-mass/faint galaxies, and cover a large dynamic range in stellar mass or rest-frame optical magnitude.
Local studies suggest potential mass-dependences of the \rline\ at $\log M_{\ast}[M_{\odot}]<8.0$ and $\log M_{\ast}[M_{\odot}]>10.0$ \citep[e.g.,][see also, Section \ref{subsec:resultcomp}]{2012ApJ...744...44W}, which is not covered in this study. 
The large dynamic range would also enable us to directly compare the sample across a wide redshift range, since the number density of high-mass galaxies decreases at high-$z$ (Figure \ref{fig:sample}).
The target lensing cluster field in this work is suitable for deep and small FoV observations, and future shallow and large FoV surveys will be complementary \citep[e.g., MINERVA,][]{2025arXiv250719706M}.

\subsubsection{Model caveats}\label{subsubsec:modcaveat}

In this work, we utilize \rline, which is measured from nebular emission lines.
The emission line strength depends on several physical parameters apart from SFH, such as the intrinsic stellar ionizing spectrum and nebular conditions.
These conditions may differ significantly between the local universe and $z\gtrsim3$; for instance, low [O\,{\sc iii}]/H$\beta$ ratios predicted by \texttt{fsps} at low $Z$ (e.g., $Z<0.15Z_{\odot}$) may not be consistent with the observations (Appendix \ref{appendix:oiiihbcomp}).

The toy model we use in this work to interpret our observations makes a huge number of simplifying assumptions. 
In particular, we assume that all galaxies in the sample are drawn from the same underlying family of SFHs, and that the simple shape of those SFHs can be characterized by a three-parameter model. 
More realistic hydrodynamical simulations suggest a much larger diversity in SFH shapes, and significantly more complex behavior \citep[e.g.,][]{2023MNRAS.526.2665S,2023MNRAS.522.3831F}. 
Furthermore, there may be more than one family of underlying SFHs in the full galaxy population.
For instance, as shown in Figure \ref{fig:popfit}, the \texttt{fsps} model with ($\alpha, \delta t, A_{\rm burst})=(0.004, 400\,{\rm Myr}, 2.0\,{\rm dex})$ cannot explain several fraction of sources with large \rha\ and small \rha, as seen in ``tail'' of the histogram.
Qualitatively, the ``tail'' seems to be extended toward both high \rha\ and low \rha\ regimes in the population models with $\delta t\lesssim200\,{\rm Myr}$ and/or $A_{\rm burst}\gtrsim2.0\,{\rm dex}$.
Because of these large simplifying assumptions, we emphasize that our toy model does not bring us an ``unique'' solution of the SFHs, and is intended to help us explore the implications of constant scatter in \rline\ across redshift. 

The model allows us to demonstrate that burstiness does not {\it need} to evolve to reproduce the observed number of UV-bright galaxies at $z>10$; however, it does not allow us to definitively quantify the ``correct'' SFH to describe high-redshift galaxies. 
We explore other possible models consistent with our observations in Appendix \ref{appendix:comp_app}.
Further testing with, for example, mock observations from simulations, will allow us to improve these models in the future to work towards a more complete picture of the SFHs of distant galaxies.

%
%
%
%
%
%
\section{Summary}\label{sec:summary}
In this paper, we investigated galaxy star formation histories (SFHs) and their evolution with both stellar mass and redshift through measurement of the line-to-UV ratio and equivalent width of both \hanii\ and \oiiihb.
We constructed mass($M_{\ast}$)- and flux($m_{\rm opt}$)-complete samples at $z=3$--$9$, avoiding any significant sample selection biases resulting from an incomplete spectroscopic targeting strategy. 
We then measured line and continuum strength using a photometro-spectroscopic dataset which includes all 20 NIRCam broad- and medium-band filters taken by the UNCOVER/MegaScience survey \citep{2024ApJ...974...92B,2024ApJ...976..101S}.
From these empirical line/continuum measurements, and through comparison with a simple population model constructed with \texttt{fsps}, we found:

\begin{itemize}
\item our photometric measurements reproduce spectroscopic measurements without any significant systematic offset. 
The normalized deviation $\Delta=({\rm NIRCam}-{\rm NIRSpec})/{\rm NIRSpec}$ of EW and \rline\ are $\leq0.05\pm0.10\,{\rm dex}$. 
We do find that the $z\sim3$ sample exhibits $\sim50\%$ larger scatter in the $\Delta$ value than the higher-redshift subsamples, potentially due to the filter coverage, metallicity variation, or increased impact of dust. 

\item The 50th percentile value of the \hanii-to-UV ratio ($P^{50}_{\text{\logrha}}$) roughly falls within the equilibrium value ($-1.93\leq\text{\logrha}\leq-1.78$) across our full redshift range. 
The 50th percentile value of the \oiiihb-to-UV ratio ($P^{50}_{\text{\logroiii}}$) also mostly falls within a range scaled by a factor of 2 from the equilibrium value of \rha; this can be explained simply by the typical line ratios H$\alpha$/H$\beta\sim3$ and [O\,{\sc iii}]/H$\beta\sim6$ at $z>3$.

\item The intrinsic scatter in the line-to-UV ratio (\ppha\ and \ppoiii) as derived from our Bayesian modeling does not depend significantly on either stellar mass or rest-frame optical magnitude, implying that SFH shapes are similar from $\log M_{\ast}[M_{\odot}]=8.0$--9.5.
These results are consistent with the local universe, where scatter does not significantly increase until $\log M_{\ast}[M_{\odot}]\lesssim8$ \citep{2019ApJ...881...71E}. 
We do not find any clear redshift dependence of the intrinsic scatter across $3<z<7$.

\item The \roiii\ distribution and its scatter shares a similar redshift trend with \rha\ at $3<z<7$, suggesting that \oiiihb\ is a statistically useful probe of \rline\ (i.e., extreme line ratios caused by alternative ionizing sources are not a majority of the sample).
This consistency allows us to extend our redshift range to $z\sim8$ with \roiii\ distributions. 
This $z\sim8$ sample shows the largest scatter among all redshift samples.

\item The comparison between \texttt{fsps} population models and \rha\ and \ewha\ distribution favors rising, long-duration, large-amplitude bursts at all redshift and mass ranges.
We obtained similar conclusions if we use \oiiihb\ instead of \hanii.
As the scatter of \rline\ is mostly affected by the amplitude of the burst ($A_{\rm burst}$) among three parameters of the \texttt{fsps} population model, the larger \roiii\ scatter in the $z\sim8$ sample may indicate a larger amplitude burst (i.e., more bursty SF) at $z>7$.

\item the population model that is most consistent with our $z\sim6$ observations $(\alpha, \delta t, A_{\rm burst})=(0.004, 400\,{\rm Myr}, 2.0\,{\rm dex})$ can boost rest-frame UV brightness by up to $\sim2\,{\rm mag}$ compared with a 200Myr constant SFH. 
This boost increases the number of UV-bright galaxies at fixed stellar mass and reproduces the observed $z\sim12$ UVLF.
This demonstrates that no increase in burstiness ($A_{\rm burst}$) is required to reproduce the observed $z\sim12$ UVLF, though larger burst amplitudes like $A_{\rm burst}\sim2.6\,{\rm dex}$ are also consistent with UVLF observations. 
Instead, model predictions for the UVLF ($A_{\rm burst}\sim0.8\,{\rm dex}$) may have underestimated the burstiness.

\end{itemize}

In combination with the great dataset taken in UNCOVER/MegaScience and empirical photometric line/continuum measurements, we reveal no significant evolution of the burstiness at $z\sim3$--7.
We also demonstrate that $z\sim3$--7 galaxies are bursty enough to reproduce $z>10$ UV LF, from the comparison with the toy model by \texttt{fsps}.
As summarized in Section \ref{subsec:caveat}, further large spectroscopic observations will overcome several caveats and improve our understanding of star formation histories at high redshift.
Expanding the redshift comparison toward $z=0$ with datasets such as COMBO-17 \citep{2003A&A...401...73W} is also important future work.

\acknowledgments
I.M. and K.A.S. thank Shelly Meyett, the MegaScience Program Coordinator, for invaluable assistance in designing WOPR 88967 and ensuring that this program was fully observed. 
This proposal was conceived of and developed at the International Space Science Institute (ISSI) in Bern, through ISSI International Team project \#562. 
D.J.S. and J.R.W. acknowledge that support for this work was provided by The Brinson Foundation through a Brinson Prize Fellowship grant.
B.W. acknowledges support provided by NASA through Hubble Fellowship grant HST-HF2-51592.001 awarded by the Space Telescope Science Institute, which is operated by the Association of Universities for Research in Astronomy, In., for NASA, under the contract NAS 5-26555.
This work is based on observations made with the NASA/ESA/CSA James Webb Space Telescope. 
The raw data were obtained from the Mikulski Archive for Space Telescopes at the Space Telescope Science Institute, which is operated by the Association of Universities for Research in Astronomy, Inc., under NASA contract NAS 5-03127 for JWST. 
These observations are associated with JWST Cycle 2 GO program \#4111, and this project has gratefully made use of a large number of public JWST programs in the A2744 field, including JWST-GO-2641, JWST-ERS-1324, JWST-DD-2756, JWST-GO-2883, JWSTGO-3538, and JWST-GO-3516. Support for program JWSTGO-4111 was provided by NASA through a grant from the Space Telescope Science Institute, which is operated by the Association of Universities for Research in Astronomy, Incorporated, under NASA contract NAS5-26555.
I.M. acknowledges funding from JWST-GO-04111.035.
K.G. and T.N. acknowledge support from Australian Research Council Laureate Fellowship FL180100060.


\appendix

\section{Results in the flux-complete samples}\label{appendix:fluxcomp}

\begin{figure*}[tb]
\begin{center}
\includegraphics[width=18cm,bb=0 0 1000 650, trim=0 1 0 0cm]{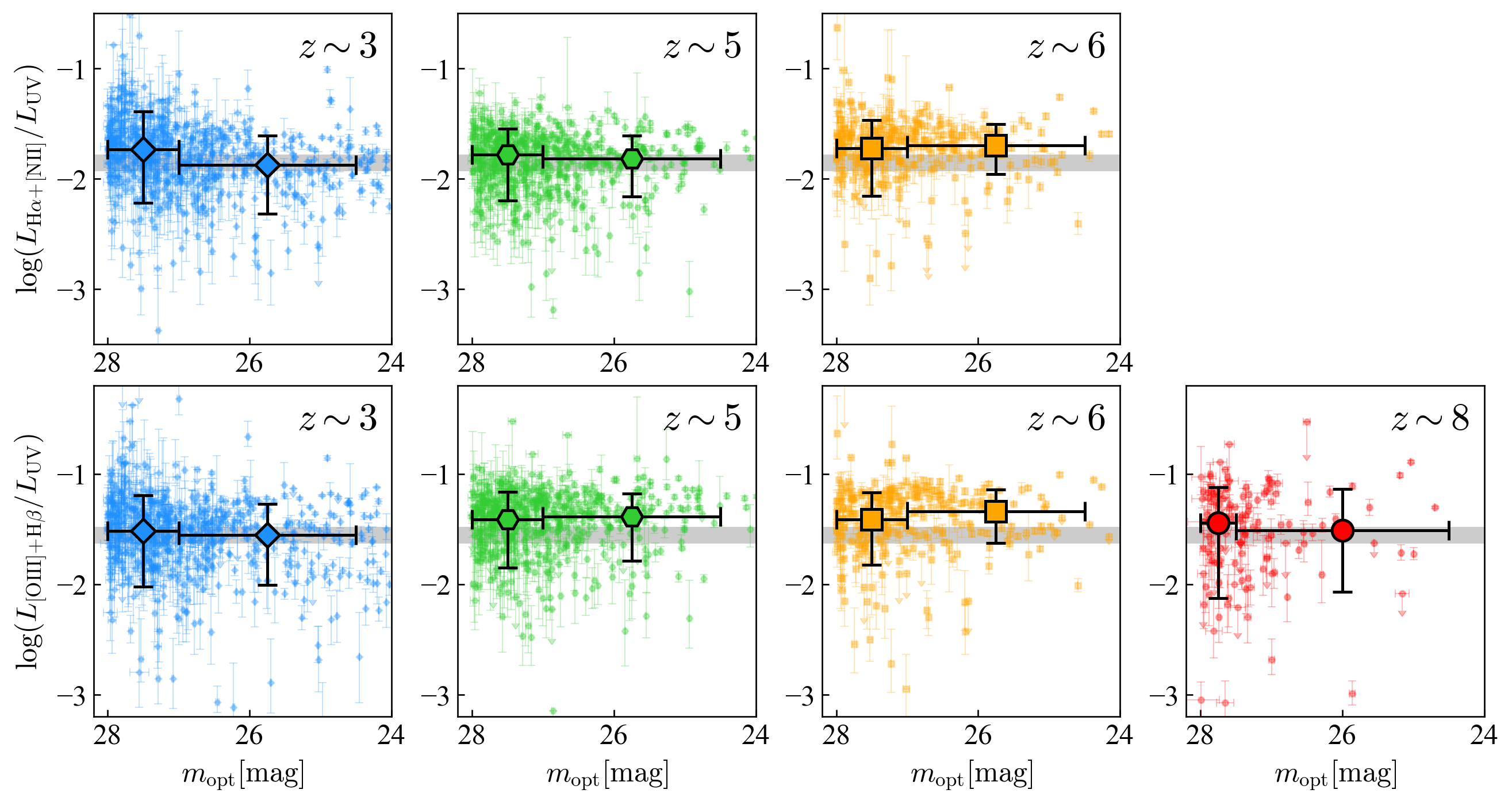}
\caption{Same with Figure \ref{fig:masscomp}, but as a function of $m_{\rm opt}$. The $z\sim3$, 5, and 6 samples flux bins are divided into $m_{\rm opt}=27$--28\,mag and 24.5--27\,mag, while $z\sim8$ sample is split into $m_{\rm opt}=27.5$--28\,mag and 24.5--27.5\,mag so that each bin contains the similar number of objects. The scatters apparently get larger at the fainter galaxies across all redshift range.}
\label{fig:fluxcomp}
\end{center}
\end{figure*}

We construct the flux-complete sample in addition to the mass-complete sample in Section \ref{subsec:sample} to check whether the stellar mass estimation from \texttt{Prospector} affects the overall distribution of \rline.
Figure \ref{fig:fluxcomp} shows the \rline\ as a function of the rest-frame $\sim4000$\AA\ magnitude.
Here, we separate the sample at each redshift into two different flux bins at $m_{\rm opt}=27\,{\rm mag}$ ($27.5\,{\rm mag}$ for $z\sim8$ sample).
The overall trend is similar to that in the mass-complete sample shown in Figure \ref{fig:masscomp}.

We find several differences between the mass-complete and flux-complete samples.
First, $P^{50}_{\text{\logrha}}$ and $P^{50}_{\text{\logroiii}}$ in the flux-complete sample are $\sim0.1\,{\rm dex}$ higher than those of the mass-complete sample.
This is likely due to the boost of the apparent $m_{\rm opt}$ from bright emission lines at the low-mass galaxies.
Because we use the broad-band covering rest-frame $\sim4500$--$7500\,\mu{\rm m}$ depending on the object's redshift to compute $m_{\rm opt}$, at the given $m_{\rm opt}$ bin, galaxies containing certain stellar masses with normal \hanii\ and \oiiihb\ emission line strength and lower-mass galaxies with strong \hanii\ and \oiiihb\ emission line are included.
Based on the usual stellar mass function \citep[e.g.,][]{2006A&A...459..745F,2017A&A...605A..70D,2023A&A...677A.184W,2024MNRAS.533.1808W}, lower-mass galaxies are more numerous than higher-mass galaxies.
Therefore, such low-mass, strong line emitters dominate the given $m_{\rm opt}$ bin and enhance the average \logrha\ and \logroiii\ in the flux-complete sample rather than the mass-complete sample.
Second, \ppha\ and \ppoiii\ seem to have stronger dependence on $m_{\rm opt}$ than $M_{\ast}$. 
This is likely due to an increase in our line and continuum measurement uncertainties for fainter galaxies. 
As shown in Figure \ref{fig:sample}, the mass-complete sample primarily includes brighter objects ($m_{\rm opt}\gtrsim27\,{\rm mag}$), while the flux-complete sample includes fainter objects ($m_{\rm opt}\gtrsim28\,{\rm mag}$).
Since our line/continuum measurement utilizes photometric observations at nearly uniform depths, fainter objects naturally have larger measurement uncertainties, increasing the scatter in the overall distribution.

\begin{figure*}[tb]
\begin{center}
\includegraphics[width=18cm,bb=0 0 1000 650, trim=0 1 0 0cm]{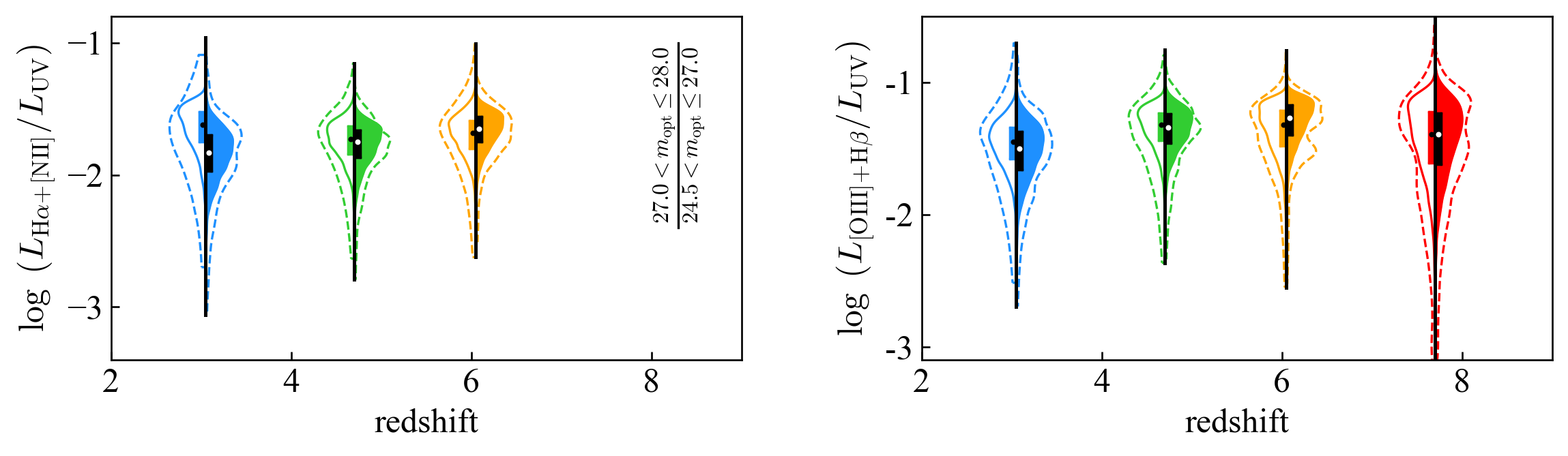}
\caption{Same with Fig \ref{fig:zevolution1}, but for the flux-complete samples. The left and right sides represent faint $m_{\rm opt}$ and bright $m_{\rm opt}$ samples, respectively.}
\label{fig:zevolution3}
\end{center}
\end{figure*}

\begin{figure*}[tb]
\begin{center}
\includegraphics[width=18cm,bb=0 0 1000 650, trim=0 1 0 0cm]{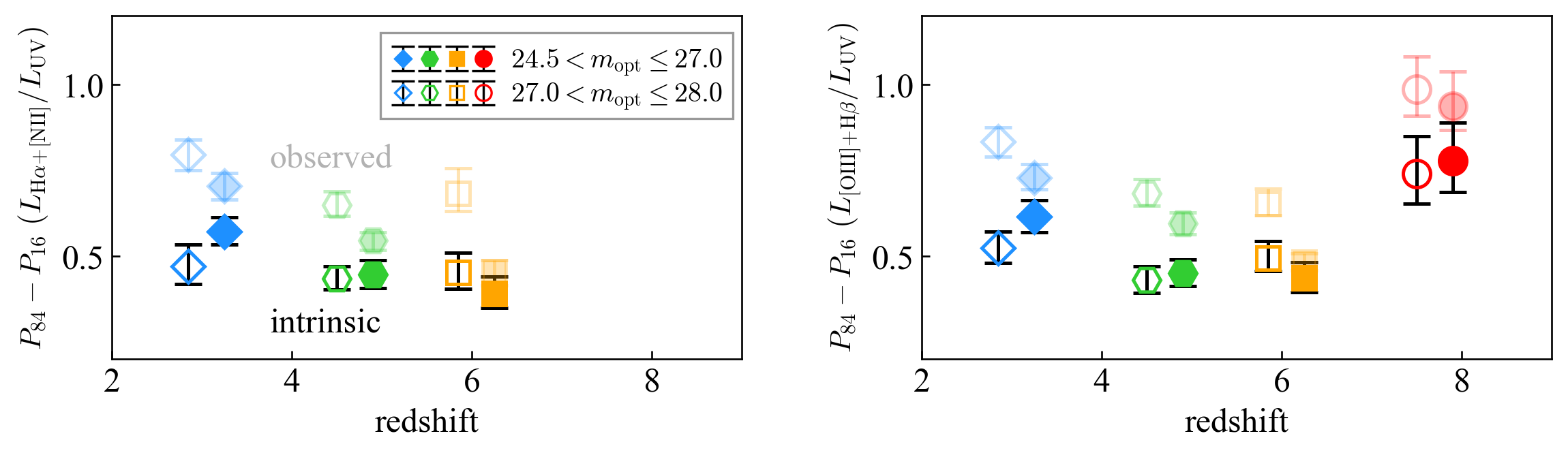}
\caption{Same with Fig \ref{fig:zevolution2}, but for the flux-complete samples. The open markers and filled markers show faint $m_{\rm opt}$ and bright $m_{\rm opt}$, respectively.}
\label{fig:zevolution4}
\end{center}
\end{figure*}

In Figure \ref{fig:zevolution3} and \ref{fig:zevolution4}, we show the intrinsic distributions and its scatter from the Bayesian population modeling of the bright ($m_{\rm opt}\leq27\,{\rm mag}$) and faint subsamples ($m_{\rm opt}>27\,{\rm mag}$) as introduced in Section \ref{subsec:zevo}.
In general, the intrinsic scatter shows no strong flux dependence as in Figure \ref{fig:zevolution1}.
As discussed in Section \ref{subsec:rline}, in the mass-complete sample, there is no significant dependence of the observed scatter on mass; however, in the flux-complete sample, we find higher observed scatter for fainter galaxies.
This flux dependence almost disappears after accounting for measurement error: the intrinsic scatters are consistent within $1\sigma$ between the bright and faint subsamples.

%
%
%
%
%
%
%
%
%
%
%
%

%
%
%
%
%
%
%
%
%
%
%
%

%
%
%
%
%
%

\section{Effect of the dust extinction correction}\label{appendix:dust_corr}

In Section \ref{subsec:photmeasure}, we correct the dust extinction based on the \texttt{Prospector} fitting.
In this section, we evaluate how the correction affect our measurements.

\begin{figure*}[tb]
\begin{center}
\includegraphics[width=18cm,bb=0 0 1000 650, trim=0 1 0 0cm]{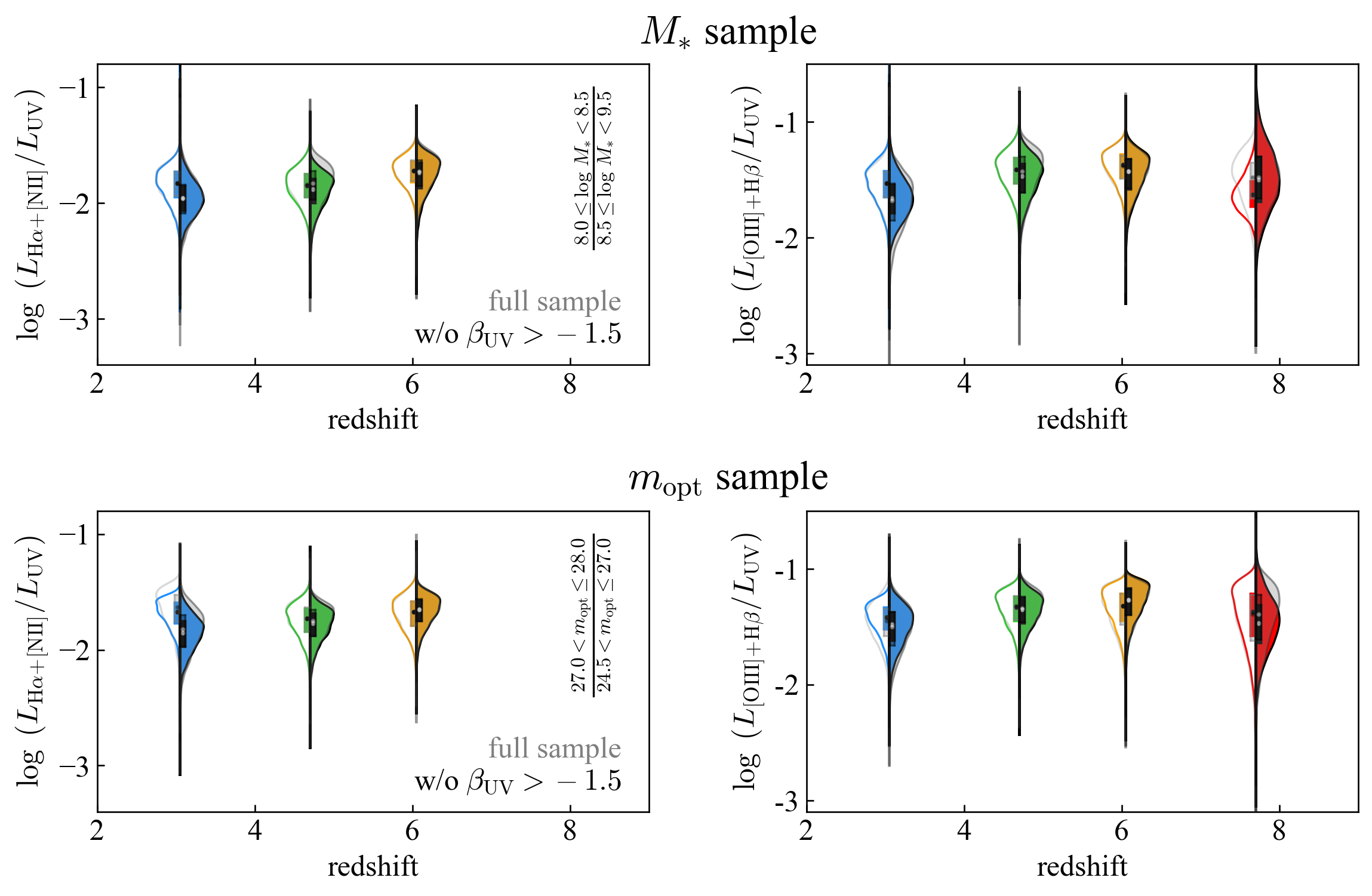}
\caption{Same with Figure \ref{fig:zevolution1}, but for the results after excluding dusty sources ($\beta_{\rm UV}>-1.5$). The results from the full sample have also been shown in transparent gray curves as a reference. There is no significant difference if we exclude the objects having relatively large dust extinction corrections, while the $z\sim8$ sample shows a slightly different distribution, potentially due to the small sample size.}
\label{fig:zevolution1_dust}
\end{center}
\end{figure*}

\begin{figure*}[tb]
\begin{center}
\includegraphics[width=18cm,bb=0 0 1000 650, trim=0 1 0 0cm]{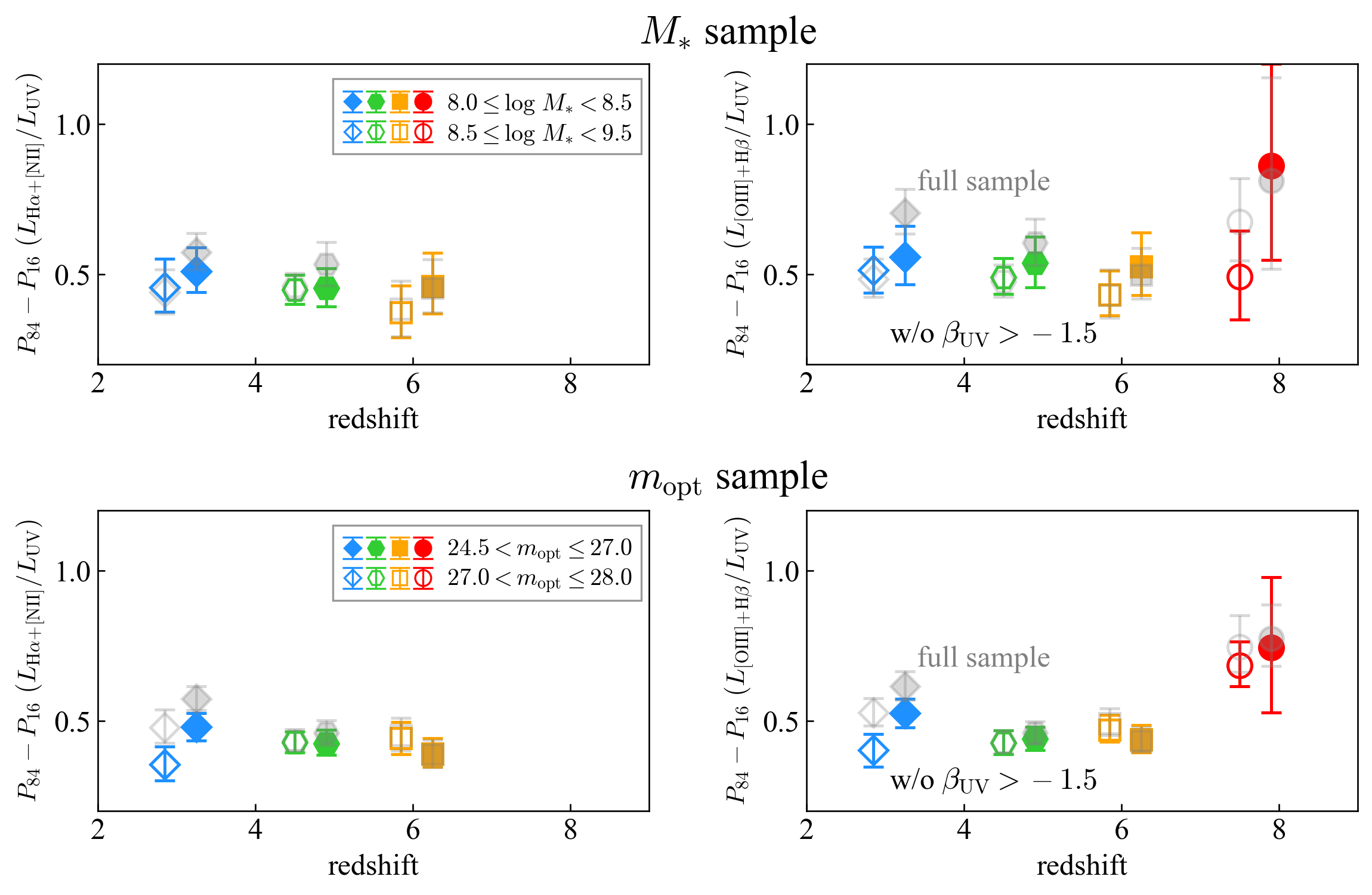}
\caption{Same with Figure \ref{fig:zevolution2}, but for the results after excluding dusty sources ($\beta_{\rm UV}>-1.5$). The results from the full sample have also been shown in transparent gray markers as a reference. There is no significant difference if we exclude the objects having relatively large dust extinction corrections, especially at $z>4$, while high-$M_{\ast}$ and/or low-$z$ samples exhibit a small difference. }
\label{fig:zevolution2_dust}
\end{center}
\end{figure*}

Figure \ref{fig:zevolution1_dust} and \ref{fig:zevolution2_dust} shows the same figure as Figure \ref{fig:zevolution1} and \ref{fig:zevolution2}, respectively, but for the results without dusty sources ($\beta_{\rm UV}>-1.5$).
Such red sources occupy $\sim20\%$ of the samples on average.
Overall, \rline\ distributions do not change significantly across the entire redshift range (Figure \ref{fig:zevolution1_dust}), suggesting that the dust extinction correction does not affect our conclusion.
The $z\sim8$ sample shows a slightly different distribution from those based on the full sample, potentially because of the small sample size.
For the mass-complete sample, the difference between the full sample and the sample without red sources is remarkable at  $z\sim3$ and 5 high-$M_{\ast}$ samples.
This implies that insufficient dust extinction correction makes a slightly uncertain measurement of the $P^{84-16}$ at such a redshift and mass range, due to significant dust extinction.
The $z\sim8$ sample also has slightly different values, potentially due to the small sample size.
For the flux-complete sample, although the $z\sim3$ sample exhibits a systematic offset, potentially due to a large dust amount at around cosmic noon, the rest of the redshift sample does not have any clear differences.

To evaluate the effect of the dust extinction correction in our SFH modeling (Section \ref{sec:results1}), we check that $A_V$ values from \texttt{Prospector} fit used to correct \rha\ and \roiii\ are consistent with those based on our bursty SFH derived in Section \ref{subsec:modelcomp}.
We start from observed \logrha, and get intrinsic, extinction-corrected \logrha\ based on the best-fit attenuation curve parameterized by $A_V$, $n$, and $f_{\hat{\tau}}$ with \texttt{Prospector}.
Then, we obtain an SED at the phase corresponding to intrinsic \logrha\ from our bursty SFH.
As the SED from the bursty SFH does not include any dust effect, we apply an attenuation effect in the SED and fit it to the original observed photometry.
The free parameters in this demonstration are $A_V$ and an arbitrary scaling factor, and we fix $n$ and $f_{\hat{\tau}}$ to the best-fit value in \texttt{Prospector} fit.
The fitting procedures are similar to Section \ref{subsec:photmeasure} - we utilize \texttt{sedpy} to convert model SED to fluxes and use \texttt{emcee} for the MCMC fitting.
If $A_V$ derived from \texttt{Prospector} (hereafter referred to as $A_V\,{\rm (Prospector)}$) and this demonstration (hereafter referred to as $A_V\,{\rm (bursty)}$) is consistent, which means extinction correction does not significantly affect the overall distribution of \logrha.

\begin{figure*}[tb]
\begin{center}
\includegraphics[width=12cm,bb=0 0 1000 650, trim=0 1 0 0cm]{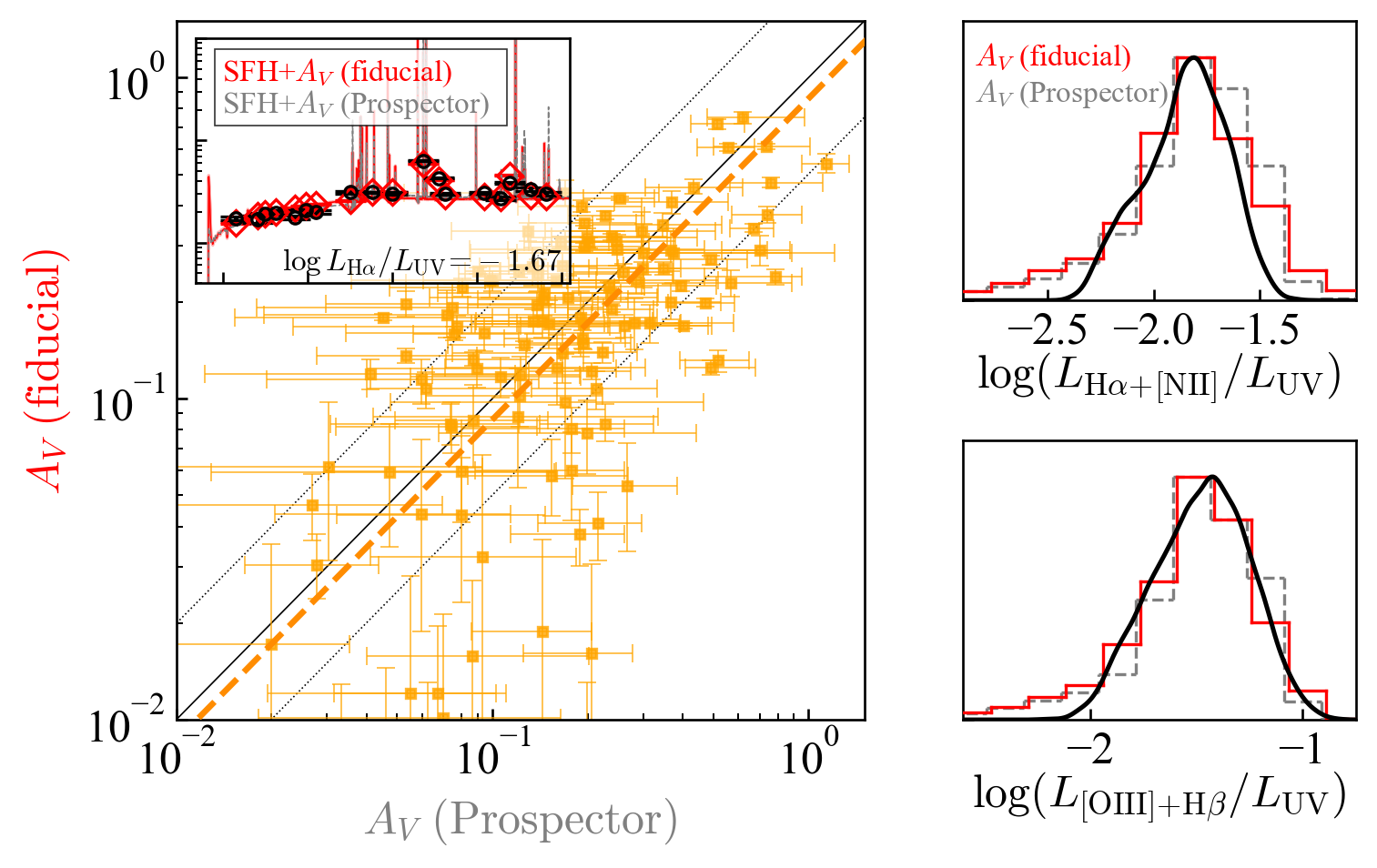}
\caption{(left) Comparison between $A_V$ from \texttt{Prospector} and that from assumption of our bursty SFH. They are broadly consistent within a factor of $\sim2$ (black dot lines), without any significant systematic offset ($\sim-0.06\,{\rm dex}$, orange dashed line). The inserted panel shows a best-fit SED from \texttt{Prospector} (gray) and an SED with our bursty SFH with best-fit $A_V$ (red) in a representative galaxy. (right) \logrha\ and \logroiii\ distribution with \texttt{Prospector}-based (gray) or our fit-based dust-correction (red). The corrected distributions do not significantly depend on attenuation correction, and have a good agreement with those from our bursty SFH (black).}
\label{fig:avcheck}
\end{center}
\end{figure*}

We compare $A_V\,{\rm (Prospector)}$ and $A_V\,{\rm (bursty)}$ in Figure \ref{fig:avcheck}.
They are pretty consistent with each other, without any significant systematic offset ($\sim-0.06\,{\rm dex}$).
We check the \logrha\ and \logroiii\ distributions based on these two different $A_V$s, and confirm that both $A_V\,{\rm (Prospector)}$-corrected and $A_V\,{\rm (bursty)}$-corrected distributions have a good agreement with those in our bursty SFH.

%
%
%
%
%
%
%
%
%
%
%
%

%
%
%
%
%
%
%
%
%
%
%
%

%
%
%
%
%
%

\section{models in \oiiihb\ emission line}\label{appendix:oiiihbcomp}
\begin{figure*}[htbp]
\begin{center}
\epsscale{1.15}
\includegraphics[width=17cm,bb=0 0 1000 650, trim=0 1 0 0cm]{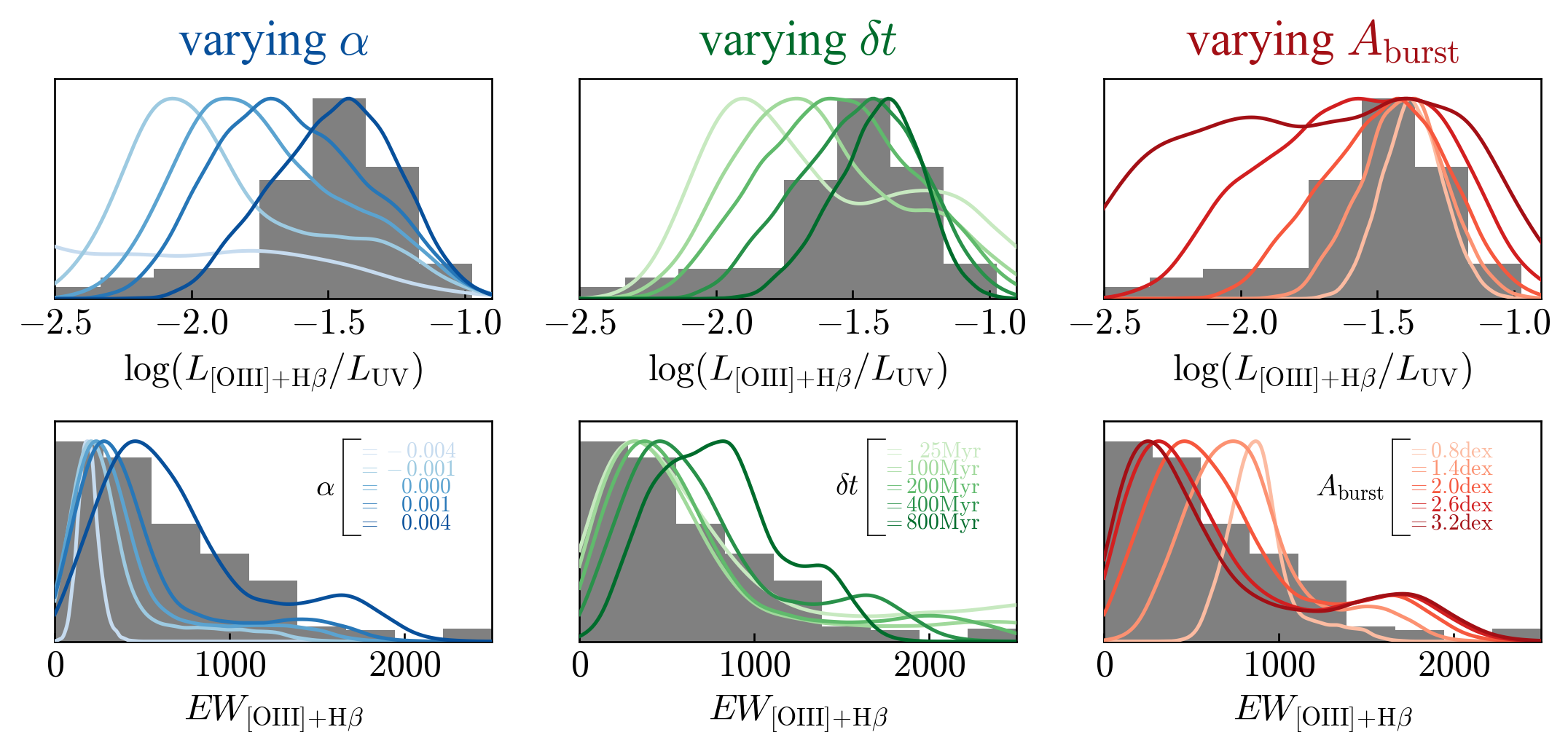}
\caption{Same with the middle and bottom panels of Figure \ref{fig:modeldemo}, but for \oiiihb\ emission lines. The overall dependence on the SFH parameters ($\alpha$, $\delta t$, $A_{\rm burst}$) is similar to the \hanii\ emission lines, suggesting \oiiihb\ is also a useful tracer of the SFH, while \oiiihb\ requires more attention from the perspective of the ionizing sources not originating from star formation, such as AGNs or SN shocks.}
\label{fig:modeldemo2}
\end{center}
\end{figure*}

In Section \ref{subsec:modelcomp}, we illustrate observed \rha\ and \ewha\ distributions and toy models with \texttt{fsps}.
Here we show the toy models with \texttt{fsps} in \roiii\ and \ewoiii, following Section \ref{subsec:model}, and compare them with the observed distribution.
In the toy model for the comparison with $[{\rm O\,III}]+{\rm H}\beta$ measurements, we adjusted the metallicities to $Z=0.16\,Z_{\odot}$ as \texttt{fsps} underestimate [O\,{\sc iii}]/H$\beta$ ratio at $Z<0.15\,Z_{\odot}$ compared with recent observational constraints \citep[e.g.,][see also \citealt{2017ApJ...840...44B} for nebular line modeling in \texttt{fsps}]{2023ApJS..269...33N,2023MNRAS.518..425C}.
The scatter due to the variation of the metallicity and ionization in ranges of $Z=0.16$--$0.5\,Z_{\odot}$ and $\log U_{\rm ion}=-2.0$--$-1.3$ is virturally reproduced by convolving a 1D Gaussian function with $\sigma_{\text{\logroiii}}=0.05$ and $\sigma_{{\rm EW},{\text{[O\,{\sc iii}]}}}=50\,$\AA into our corresponding \texttt{fsps} model distributions.
In Figure \ref{fig:modeldemo2}, we compare our \texttt{fsps} models for \oiiihb\ emission lines.
The burst models for \oiiihb\ line follow a similar dependence on the SFH parameters as in Figure \ref{fig:modeldemo}.
As far as galaxies within the sample share similar metallicities and ionizing parameters within ranges of $Z=0.2$--$0.5\,Z_{\odot}$ and $\log U_{\rm ion}=-2.0$--$-1.3$ (see Section \ref{subsec:model}), \oiiihb\ can also be a good tracer of the SFH.
Note that \oiiihb\ needs more caution from the perspective of the ionizing sources not originating from star formation, such as AGNs or SN shocks.

%
%
%
%
%
%

\section{the comparison of the observed distributions and \texttt{fsps} model in EW distribution}\label{appendix:comp_app}

In this section, we describe the details about the comparison between the observed \rline\ distributions and those from our toy model SFH.
As shown in Figure \ref{fig:popfit}, the KS and AD test support the three models with ($\alpha, \delta t, A_{\rm burst})=(0.004, 400\,{\rm Myr}, 2.0\,{\rm dex})$, ($0.004$, 800\,Myr, 2.6\,dex) and ($0.004$, 800\,Myr, 3.2\,dex) can reproduce the observed \rline\ distributions.

\begin{figure*}[tb]
\begin{center}
\includegraphics[width=15cm,bb=0 0 1000 650, trim=0 1 0 0cm]{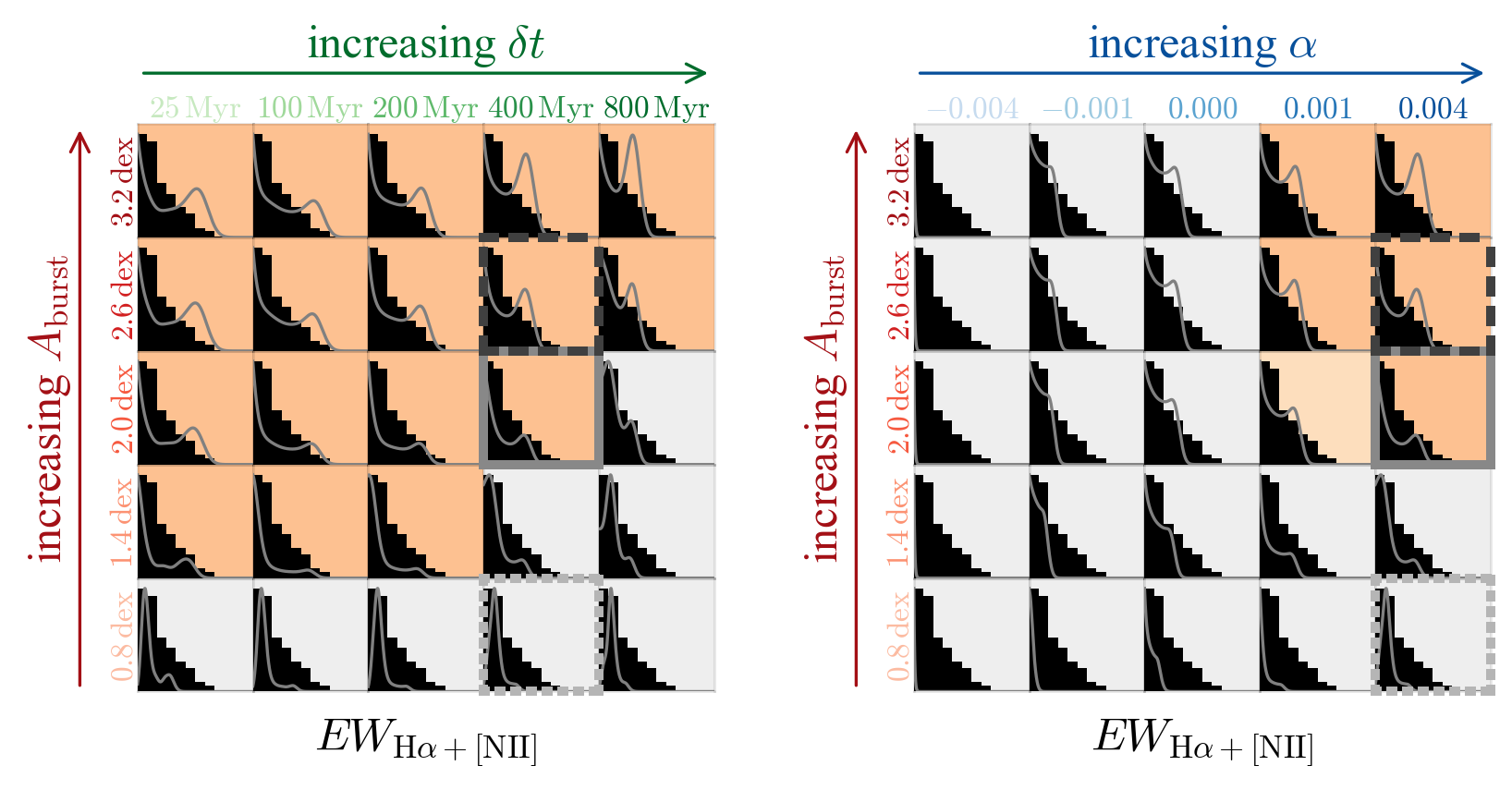}
\caption{The same figure with Figure \ref{fig:popfit}, but for \ewha\ distribution. 
The plausible models inferred from \ewha\ distribution completely include those from \rha, suggesting weak constraining power compared with \rha.}
\label{fig:popfit2}
\end{center}
\end{figure*}

Figure \ref{fig:popfit2} shows the comparison in EW, instead of \rline\ in Figure \ref{fig:popfit}.
A much broader range of models is able to reproduce the observed EW distributions, indicating that \rline\ has more constraining power on the SFH than EW does.
We expect that this is due to the wavelength difference, which is equivalent to the sensitivity to the star formation.
While both \rline\ and EW are fundamentally the ratio between emission line and continuum flux, \rline\ uses the rest-frame UV continuum in the denominator while EW uses the rest-frame optical continuum.
In general, the rest-frame UV continuum traces the SF activity, whereas the optical continuum traces the existing stellar mass.
\rline\ reflects SFH variations more than EW, and therefore has a large dynamic range, especially for faint line emitters.
Indeed, galaxies that occupy the $R({\rm line})\lesssim-1.85\,(={\rm equilibrium\ value})$ regime are distributed at $\lesssim500\,\text{\AA}$ in the EW.
Even if a population-modeled distribution shows only too faint emitters in \rline, for EW, most of them are concentrated in $\lesssim500\,\text{\AA}$ range (see green thin lines in the middle panel of Figure \ref{fig:modeldemo}), so most of the population models are classified as not statistically bad.
That suggests an advantage to use \rline\ to constrain SFHs, even though it requires taking more care about dust extinction than EW (see Appendix \ref{appendix:dust_corr}).

%
%
%
%
%
%

\section{No burst models}\label{appendix:noburst_models}

\begin{figure*}[tb]
\begin{center}
\includegraphics[width=12.0cm,bb=0 0 1000 650, trim=0 1 0 0cm]{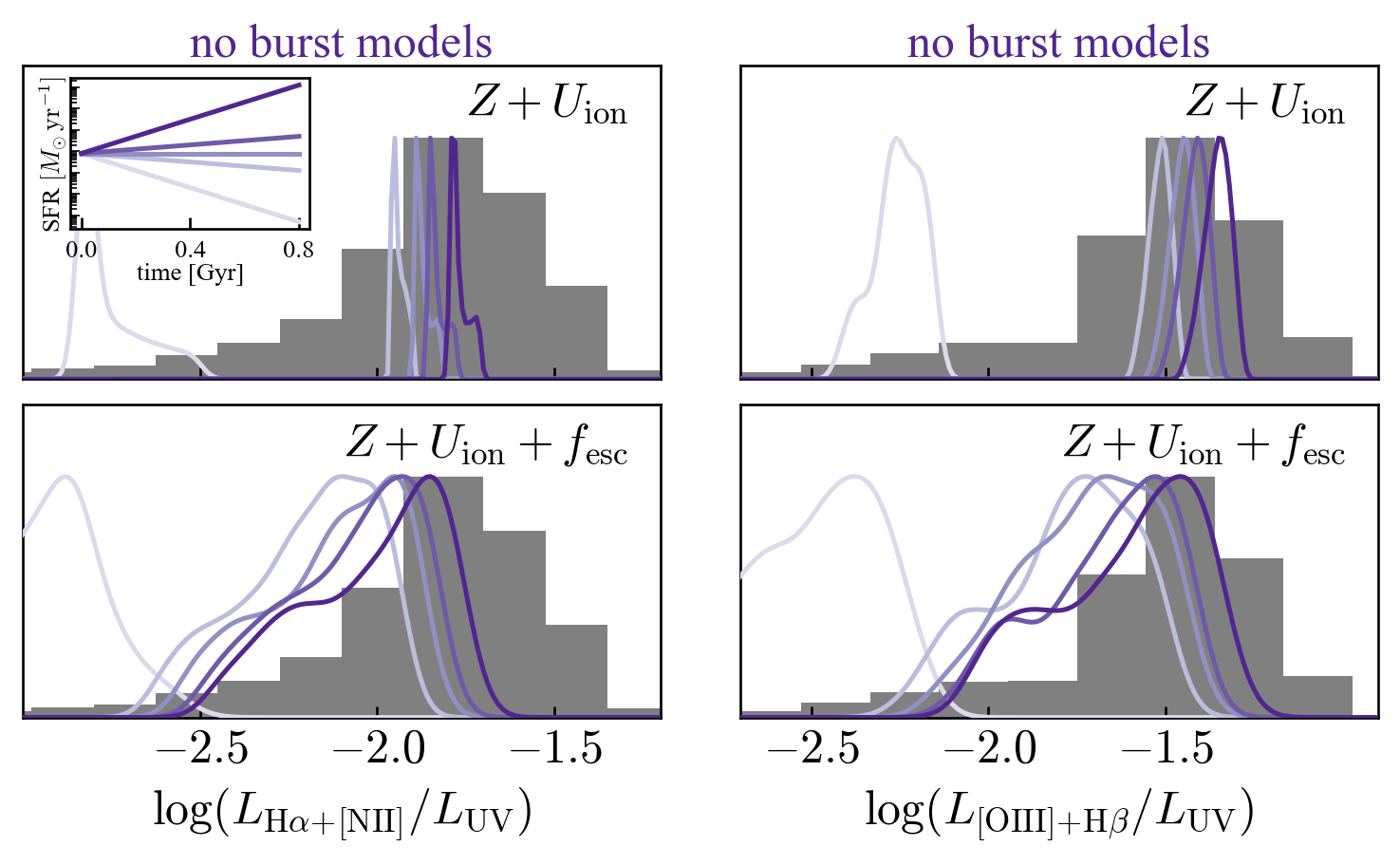}
\caption{Same with Figures \ref{fig:modeldemo} and \ref{fig:modeldemo2}, but overlayed by the population models without bursts. Only \rha\ and \roiii\ distributions are shown, given that \rha\ or \roiii bring stronger constraints than \ewha\ or \ewoiii\ (Figure \ref{fig:popfit}). The top panel shows the model with the variation in metallicity $Z$ and ionization parameter $U_{\rm ion}$, and the bottom panel illustrates the model with the variation in $Z$, $U_{\rm ion}$, and escape fraction of the ionizing photon $f_{\rm esc}$. The population models without bursts are not sufficient to explain the overall observed distribution.}
\label{fig:badmodelcomp}
\end{center}
\end{figure*}

To check whether the variation in EWs and \rline\ can be explained without bursty SFHs, we compare our observations to the population models without bursts created in Section \ref{subsec:model}.
Figure \ref{fig:badmodelcomp} shows the comparison between burst-free models and our measured \rline\ and EWs distributions.
Here, the variation of metallicity $Z$ and the ionizing parameter $U_{\rm ion}$ has been included in the \texttt{fsps} calculation (the top panel in Figure \ref{fig:badmodelcomp}, hereafter referred to as $Z+U_{\rm ion}$ models).
We also compare the models with an additional parameter, escape fraction of the ionizing photon ($f_{\rm esc}$), which changes EWs and \rline, in the bottom panel (hereafter referred to as $Z+U_{\rm ion}+f_{\rm esc}$ models).
Here we simply scaled the EWs and $R({\rm lines})$ by $(1-f_{\rm esc})$, in a randomly selected $f_{\rm esc}$ value from a range of $f_{\rm esc}=0$--0.8.
Note that this $f_{\rm esc}$ range is much more extended than actual observational constraints \citep[typical $f_{\rm esc}\sim0.1$ at $z\sim4$--9, e.g.,][]{2023A&A...672A.155M,2024A&A...685A...3M}, to show the most extreme case.
The effect of the $f_{\rm esc}$ results in an extension toward the low EW/\rline\ regime.
The maximum $p$-values in \ewha, \logrha, \ewoiii, and \logroiii\ distributions from $Z+U_{\rm ion}$ or $Z+U_{\rm ion}+f_{\rm esc}$ models are only $p_{\rm KS}=0.004$ and $p_{\rm AD}=0.001$, suggesting that no burst-free models can reproduce the observed distribution.
The rapid upturn of the SFR in $\sim10\,{\rm Myr}$ timescale produces the enhanced \rline\ from the equilibrium values, where $\sim50\%$ of the galaxies in the sample have such enhanced \rline.
This high fraction of the enhanced \rline\ galaxies directly suggests that the burst is likely to be necessary.
We found a similar conclusion in \roiii\ distributions.
This is also discussed in recent studies focusing on the extreme emission line galaxies \citep[e.g.,][]{2024MNRAS.535.1796B}, nappers/mini-quenched galaxies \citep[e.g.,][]{2025A&A...697A..88L}, star-forming main sequence \citep[SFMS, e.g.,][]{2024ApJ...977..133C,2025arXiv250804410S} and so on \citep[e.g.,][]{2024MNRAS.52711372A}.

Note that here we merely suggest it is challenging to explain the distribution using a single common slope SFH, even with dramatic variations in $f_{\rm esc}$.
As each object can have its unique slope, it may be possible to reproduce the distribution by superimposing models with different slopes.
However, the superimposition of models with different slopes is essentially equivalent to a bursty model conceptually. 
Therefore, the model with bursts, which can reproduce the entire distribution more simply, is considered more plausible.

%
%
%
%
%
%

\clearpage
\bibliography{main.bib}{}
\bibliographystyle{apj}

\end{document}